
\magnification=1200
\hsize=13cm
\centerline{\bf 1. Introduction}
\vskip 0.7cm
The study of systems of strongly correlated
electrons has received an increasing attention
in the last years, especially after the discoveries
of the Fractional Quantum Hall Effect  and the High
Temperature Superconductivity. In this context, one
of the most studied is the Hubbard model [1] which
describes the dynamics of non-relativistic electrons
moving on a lattice. Its Hamiltonian contains a kinetic
term and a quartic potential term which effectively describes
the Coulomb interactions among electrons.

Another lattice model that has been suggested as an
appropriate starting point for the theory of the High
Temperature Superconductivity [2,3], is the so-called
$t$-$J$ model [4] which is characterized by the absence
of doubly occupied electron states. Its Hamiltonian
contains a kinetic term of strength $t$ responsible
for the hopping of electrons from one lattice site to
its nearest neighbors, and a potential term of strength
$J$ which describes nearest neighbor spin exchange
interactions. If $J<\!<t$, the $t$-$J$ model can be
mapped onto a Hubbard model with very strong Coulomb
repulsion; for other values of the coupling constants
instead, the two models exhibit a quite different
behavior. A particularly interesting feature of the
$t$-$J$ model is the presence of a supersymmetry [5]:
In fact, when $J=\pm2\,t$, the Hamiltonian of the $t$-$J$
model commutes with the generators of the superalgebra
$SU(1|2)$ [6]. It is remarkable that precisely at the
supersymmetric point $J=\pm2\,t$, the $t$-$J$ model in
one dimension is exactly solvable by Bethe {\it Ansatz}
[7,8]. We recall that the supersymmetric $t$-$J$ model
cannot be derived from the Hubbard model, but instead
can be mapped onto a quantum lattice gas of hard core
bosons and fermions. This latter system was formulated
in the mid-seventies and solved exactly by Bethe {\it Ansatz}
in one dimension [9,8].

The availability of a complete solution for the one-dimensional
supersymmetric $t$-$J$ model allows a detailed analysis of
its collective excitations [10]. In particular two kinds
of low lying quasiparticle excitations have been
identified above the antiferromagnetic ground state
($J>0$): One carries charge and no spin and is called
holon; the other carries spin and no charge and is
called spinon. This remarkable phenomenon of spin-charge
separation, which is rigorously proven only in one dimension,
has been conjectured to occur also in two dimensions [3,11].
To implement this separation between spin and charge, one
often uses the formalism of the slave operators [12] and
makes the {\it Ansatz} in which the electron is a product
of a charged spinless antiholon and a neutral spinon of
spin 1/2. Since the electron is a fermion, holons and
spinons must have complementary statistics. For example,
the holon may be fermionic and the spinon bosonic (slave
fermion representation [13]), or vice-versa the holon
may be bosonic and the spinon fermionic (slave boson
representation [3,11]). Actually, since in two
dimensions the statistics is arbitrary [14,15], there
are in principle other possibilities; in particular
it has been suggested [16] that holons and spinons
are semions, {\it i.e.} anyons that are exactly half-way
between bosons and fermions.
In [17,18] it has been shown that these different
statistics assignments can be interpreted in a unified
way within the framework of the bosonization procedure
of two-dimensional systems. More precisely, the slave
boson and slave fermion representations arise as two
different gauge fixings in the abelian bosonization of
the $t$-$J$ model, whereas the semion case naturally
occurs in the non-abelian bosonization.

In this paper we present a generalization of the
analysis of [17,18], and in particular we show that
performing an abelian bosonization with two independent
Chern-Simons fields, the holons and the spinons of the
two-dimensional $t$-$J$ model may become anyons of
arbitrary complementary statistics, in such a way
that the product of one antiholon and one spinon
({\it i.e.} the electron) is a fermion. We call
this the slave anyon representation, which interpolates
continuously between the slave boson and the slave
fermion representations; in this way the semionic
statistics of [16] simply becomes a particular case.
Therefore, we deduce that at the formal kinematical
level, all consistent statistics assignments are
possible and only a {\it dynamical} analysis can
eventually select one among these many possibilities.

Another interesting fact is that the supersymmetry of the
$t$-$J$ model at $J=\pm2\,t$ becomes simpler and more
transparent when formulated in terms of holon and spinon
operators. In fact, since these have complementary statistics,
it is natural to expect that the fermionic supercharges
transform holons into spinons and vice-versa
\footnote{$^{1}$}{To avoid confusions, we stress that here
the terms holons and spinons are used in a different way
with respect to [10] where holons and spinons denote
{\it different} supermultiplets of $SU(1|2)$.}. Indeed
this is what happens, because the superalgebra $SU(1|2)$
is {\it linearly} realized on holons and spinons in contrast
to the non-linear realization in terms of the electron
operators. In this paper we discuss in particular the
supersymmetry of the $t$-$J$ model in the slave anyon
representation and analyze the braiding properties of
the supercharge densities with holons and spinons.

The paper is organized as follows: In Section 2 we review
the main features of the $t$-$J$ model, write its
Hamiltonian and present its bosonization with two
independent Chern-Simons fields to yield holons and
spinons with complementary statistics. In Section 3 we
discuss the supersymmetry of the $t$-$J$ model in the
slave fermion formalism, whereas in Section 4 we introduce
slave holons and spinons in the anyon representation and
discuss their braiding properties. Then we prove the
invariance under $SU(1|2)$ of the Hamiltonian of the
$t$-$J$ model at the supersymmetric point $J=2\,t$ in the
slave anyon representation. Finally, Section 5  presents
our conclusions, and Appendix A
contains an alternative proof of the supersymmetry
of the Hamiltonian of the $t$-$J$ model when $J=2\,t$.

\vskip 2cm
\centerline{\bf 2. The $t$-$J$ Model and its
Bosonizations}
\vskip 0.7cm
The (two-dimensional) $t$-$J$ model describes a
lattice system of strongly correlated non-relativistic
electrons. It was proposed a few years ago by Zhang
and Rice [4] as an effective
model for the electronic dynamics
in the copper-oxide planes of the high temperature
superconductors.
Given a two-dimensional square lattice $\Omega$ of
spacing $a=1$, one assumes that in each site at most
one electron can be accomodated, {\it i.e.} doubly
occupied electron states are excluded
from the physical Hilbert space of the system; this
is a way to simulate strong
Coulomb repulsion. The $t$-$J$ model with chemical
potential $\mu$
is described by the following Hamiltonian
$$
\eqalign{
H_{tJ} =~& {\cal P} \left[
-t\sum_{<i,j>}\sum_{\alpha=\pm1}
\Big(c_\alpha^\dagger(i)\,c_\alpha(j)+
{\rm h.\,c.}\Big)
\right. \cr
&~~~~\left.+J\sum_{<i,j>}\left({\bf S}(i)
\cdot{\bf S}(j)
-{1\over 4}n(i)\,n(j)\right)+\mu
\sum_{i\in \Omega}n(i)\right]
{\cal P} }
\eqno(2.1)
$$
where $\sum\limits_{<i,j>}$ denotes the sum
over nearest-neighbor lattice sites;
$c_\alpha^\dagger(i)$ and $c_\alpha(i)$ are
respectively creation and annihilation operators
for electrons with third component of the spin
$\alpha/2$ at the site $i$.
They satisfy the usual anticommutation relations
$$
\left\{c_\alpha(i)\,,\,c_\beta^\dagger(j)
\right\}=\delta_{\alpha \beta}\,\delta(i,j)
\eqno(2.2)
$$
where $\delta(i,j)$ is the lattice delta
function. The vector
${\bf S}(i)$ is the
spin operator whose components are
$$
S^a(i) = \sum_{\alpha,\beta=\pm1}
c_\alpha^\dagger(i)
\left({{\sigma^a}\over 2}\right)_{\alpha\beta}
c_\beta(i)
\hskip 1cm a=1,2,3,
\eqno(2.3{\rm a})
$$
($\sigma^a$ being the Pauli matrices), and $n(i)$ is
the number operator
$$
n(i) = \sum_{\alpha=\pm1} n_\alpha(i) =\sum_{\alpha=\pm1}
c_\alpha^\dagger(i)\,c_\alpha(i)\ \ .
\eqno(2.3{\rm b})
$$
Finally, ${\cal P}$ denotes the Gutzwiller projection
operator which ensures that
doubly occupied electrons states are removed from the
Hilbert space of the system [19].
The parameters $t$ and $J$ in (2.1) are coupling
constants which we take to be positive: $t$ denotes
the strength of the kinetic nearest-neighbor hopping term;
$J$ denotes the strength of the spin exchange
nearest-neighbor
interaction. We recall that
for $J<\!<t$ the hamiltonian (2.1) can be derived
from the Hubbard model in the limit
of (infinite) strong on-site repulsion and that for
$t=0$ it reduces to the antiferromagnetic Heisenberg XXX model [1].

In the following we will discuss the bosonization
procedure with Chern-Simons fields and the presence
of anyons, which are specific features of the
{\it two-dimensional} $t$-$J$ model; however,
most of the considerations that we present
hereinafter are valid for lattices of any dimensions.
As we mentioned, doubly occupied states are
excluded from the Hilbert space of
the $t$-$J$ model; therefore if we denote by
$|0\rangle$ the Fock vacuum which satisfies
$c_\alpha(i) |0\rangle=0$ for all $i\in \Omega$ and
$\alpha=\pm1$, there are
only three possible electronic states at any given
lattice site $i$, namely
$$
|0\rangle~~~,~~~c_+^\dagger(i) |0\rangle~~~,~~~
c_-^\dagger(i) |0\rangle\ \ .
\eqno(2.4)
$$
The fourth state $c_+^\dagger(i)\,c_-^\dagger(i)
|0\rangle$, which would be allowed
by Fermi statistics, is removed by the Gutzwiller
projector ${\cal P}$. There is an
equivalent way to realize this truncation in the
Fock space: Instead of using the operators $c_\alpha^\dagger$
to construct the physical states and then the Gutzwiller
projector to remove those
with double occupancy, one can use the {\it projected}
fermionic operators
$$
\eqalign{
{c'}_\alpha(i) &\equiv \Big(1-n_{-\alpha}
(i)\Big)\,c_\alpha(i) \ \ , \cr
{c'}_\alpha^\dagger(i) &\equiv c_\alpha^\dagger(i)
\,\Big(1-n_{-\alpha}(i)\Big)
}\eqno(2.5)
$$
for $\alpha=\pm1$. It is immediate to verify that
these operator satisfy a hard-core
constraint even when their spin index is different.
Indeed one has
$$
{c'}_\alpha(i)~{c'}_\beta(i)=
{c'}_\alpha^\dagger(i)~
{c'}_\beta^\dagger(i)=0
\eqno(2.6)
$$
for {\it all} $\alpha$ and $\beta$. Furthermore,
using (2.2) it is straightforward to verify that
$$
\eqalign{
\Big\{{c'}_\alpha(i)\,,\,{c'}_\beta(j)\Big\} &=
\Big\{{c'}_\alpha^\dagger(i)\,,\,
{c'}_\beta^\dagger(j)\Big\} =0 \ \ , \cr
\Big\{{c'}_\alpha(i)\,,\,
{c'}_\beta^\dagger(j)\Big\} &=
\delta(i,j) \left[ \left(1-{1\over 2}n(i)
\right)\delta_{\alpha\beta}+
\sum_{a=1}^3S^a(i)\sigma^a_{\alpha\beta}
\right]\ \ .
}\eqno(2.7)
$$
The operators $S^a(i)$ and $n(i)$ are defined
respectively in (2.3a) and (2.3b), but it is easy
to realize that they can be written indifferently
in terms of $c_\alpha$ or ${c'}_\alpha$. In fact,
with simple algebra we see that
$$
{S'}^a(i)\equiv \sum_{\alpha,\beta=
\pm1}{c'}_\alpha^\dagger(i)
\left({{\sigma^a}\over 2}\right)_{\alpha\beta}{c'}_\beta(i)=
\sum_{\alpha,\beta=\pm1}c_\alpha^\dagger(i)
\left({{\sigma^a}\over 2}\right)_{\alpha\beta}c_\beta(i)
=S^a(i)\ \ ,
\eqno(2.8{\rm a})
$$
and
$$
{n'}_\alpha(i)\equiv {c'}_\alpha^\dagger(i)
\,{c'}_\alpha(i) = c_\alpha^\dagger(i)
\,c_\alpha(i)\,\Big(1-n_{-\alpha}(i)\Big)^2=n_\alpha(i)\ \ ,
\eqno(2.8{\rm b})
$$
where the last equality follows from
$n_\alpha(i)\,n_{-\alpha}(i)=0$ which is a
consequence of the absence of doubly occupied
states.
Using the projected operators (2.5), the
Hamiltonian (2.1) becomes
$$
\eqalign{
H_{tJ} =& \Bigg[ -\,t\sum_{<i,j>}
\sum_{\alpha=\pm1} \Big(
{c'}^\dagger_\alpha(i)\,c'_\alpha(j) +
{\rm h.c.}\Big) \cr
&~~~+\,J\sum_{<i,j>}\Big({\bf S}(i)\cdot
{\bf S}(j) -{1\over 4}\,n(i)\,n(j)\Big)
+\mu\sum_{i\in\Omega}n(i)\Bigg]\ \ .
}
\eqno(2.9)
$$

Since the Hilbert space of the $t$-$J$ model
does not contain any doubly occupied states,
the hopping term of the Hamiltonian can be dropped
at half-filling, {\it i.e.} when the
number of electrons is equal to the number of
lattice sites. In this case
the $t$-$J$ model becomes equivalent to the
(antiferromagnetic) XXX model. The situation is
clearly different below half-filling: In such a
case there are some empty points and some electrons
can hop into them from neighboring occupied sites
with a rate that is proportional to the coupling
constant $t$. A deviation from half-filling is usually
called ``doping'' and can be realized by suitably choosing
the value of the chemical potential $\mu$ in the Hamiltonian.
Several authors suggested that, at small doping, the spectrum
of the lowest lying excitations above the antiferromagnetic
ground state consists of holons and
spinons [2,3,11]. The holons carry charge and no spin,
whilst the spinons carry spin and no
charge. Only an accurate study of the dynamics could allow
to check if
this mechanism of spin-charge separation actually occurs.
To this aim it may be useful to perform a change of variables
and write the projected electron operators, which obviously
have both charge and spin, as products of an antiholon and
a spinon. More precisely one writes
$$
{c'}_\alpha(i)= e^\dagger(i)\,s_\alpha(i)~~~~,~~~~
{c'}_\alpha^\dagger(i)
=e(i)\,s_\alpha^\dagger(i)
\eqno(2.10)
$$
where $e^\dagger(i)$ and $e(i)$ are respectively the
creation and annihilation
operators for charged spinless holons, and $s_\alpha^\dagger(i)$
and $s_\alpha(i)$ are respectively the creation
and annihilation operators for neutral spinons of
spin $\alpha$, and are such that
$$
e^\dagger(i)|0\rangle = s_\alpha(i)|0\rangle = 0
$$
where $|0\rangle$ is the Fock vacuum of Eq. (2.4)
representing a hole.
The requirement that there are no doubly occupied electron
states, translates into the constraint
$$
\phi(i)\equiv e^\dagger(i)\,e(i)\,+\sum_{\alpha=\pm1}
s_\alpha^\dagger(i)
\,s_\alpha(i)-1=0
\eqno(2.11)
$$
for all $i\in \Omega$.
This constraint implies that in each lattice site one of
the following three possibilities must occur:
Either there is one hole or there is one electron with
spin $+1$ or one electron with spin $-1$. There
are no other possibilities, just like in the original
electronic description where there were only three allowed
configurations at each site.

Another important issue is the statistics of holons and
spinons. Since the electron operators must be fermionic,
$e(i)$ and $s_\alpha(i)$ cannot have the same statistics.
In principle we have two possibilities: The holon is a
fermion and the spinon is a boson, or vice-versa, the holon
is a boson and the spinon is a fermion. These two scenarios
are usually called the slave-fermion and the slave-boson
descriptions respectively, and both of them have been
proposed and studied in the literature
(the slave-fermion approach for the $t$-$J$ model has
been considered in [13,20], while the slave-boson approach
has been considered in [3,11,21]). However, for {\it two-dimensional}
systems there
are other possibilities, since in two dimensions the
statistics is not necessarily bosonic or fermionic but
can be anyonic [22] (for a review see for instance [14,15]).
Indeed, in [16] Laughlin conjectured that the
two-dimensional holons and spinons are semions, {\it i.e.}
anyons that are exactly half-way between
bosons and fermions. Actually, as we will show later,
holons and spinons for the $t$-$J$ model in two dimensions
can in principle be anyons of arbitrary statistics with
the only constraint that
the composite object formed by one holon and one
spinon be a fermion.

A few remarks are in order at this point. First of all,
we should keep in mind that the replacement of the
electron operators
with products of holons and spinons according to
(2.10) is a purely
{\it formal} operation.
Differently from the electron operators $c'_\alpha(i)$
and $c'^\dagger_\alpha(i)$, the holon and spinon
oscillators do not commute with the constraint
operator $\phi(i)$; therefore the spin-charge
separation, {\it i.e.} the excitation of a single
holon or spinon, can occur only if the constraint
(2.11) is somehow released.
Secondly, the issue of
the statistics of holons and spinons cannot be resolved
at this formal stage, since all the above mentioned
assignments of statistics are formally correct.
In fact, to make any progress it would be necessary
to determine what is the dynamics of holons and spinons
and to find their Hamiltonian. The first step towards
this aim is simply to substitute (2.10) into the
Hamiltonian of the $t$-$J$ model.
For definiteness we choose the slave fermion description,
 but similar considerations would apply equally well to
the slave boson formalism. In the slave fermion description,
the fermionic holon operators obey
$$
\Big\{e(i)\,,\,e^\dagger(j)\Big\} = \delta(i,j)\ \ ,
\eqno(2.12{\rm a})
$$
and the bosonic spinon operators obey
$$
\Big[s_\alpha(i)\,,\,s_\beta^\dagger(j)\Big] =
\delta(i,j)~\delta_{\alpha\beta}\ \ ,
\eqno(2.12{\rm b})
$$
with all other possible (anti)commutators vanishing. A straightforward
substitution
of (2.10) into the Hamiltonian (2.9) leads to
\footnote{$^2$}{To simplify the notation, here and
in the following we understand the summation symbols
over repeated spin indices.}
$$
\eqalign{
H_{tJ} =&  \Bigg\{
+\,t\sum_{<i,j>}
\Big(e^\dagger(j)\,e(i)\,s_\alpha^\dagger(i)\,
s_\alpha(j)+{\rm h.\,c.}\Big)
 +{J\over 4}\sum_{<i,j>}e(i)\,e^\dagger(i)\,e(j)
\,e^\dagger(j)
{}~\times \cr
&~~~\times\left[\sum_{a=1}^3\Big(s_\alpha^\dagger(i)\,
\sigma^a_{\alpha\beta}\,s_\beta(i)\,
s_{\alpha '}^\dagger(j)\,
\sigma^a_{\alpha ' \beta '}\,s_{\beta '}(j)\Big)
-s_\alpha^\dagger(i)\,s_\alpha(i)\,
s_{\alpha '}^\dagger(j)\,s_{\alpha '}(j)\right]
\cr
&~~~+\sum_{i\in \Omega}
\Big[\mu~e(i)\,e^\dagger(i)\,s_\alpha^\dagger(i)\,
s_\alpha(i)+{\rm i}\, \Lambda(i)\,
\phi(i)\Big]\Bigg\}}
\eqno(2.13)
$$
where we have introduced the Lagrange multiplier
$\Lambda(i)$ to enforce the constraint (2.11).

Using the anticommutation relations of the holons,
the $J$-term of (2.13) can be rewritten as follows
$$
\eqalign{
{J\over 4}&\sum_{<i,j>}\Big[1-e^\dagger(i)\,e(i)-
e^\dagger(j)\,e(j)
+e^\dagger(i)\,e(i)\,e^\dagger(j)\,e(j)
\Big]\times \cr
&~~~\times\left[\sum_{a=1}^3\Big(s_\alpha^\dagger(i)\,
\sigma^a_{\alpha\beta}\,s_\beta(i)\,
s_{\alpha '}^\dagger(j)\,
\sigma^a_{\alpha ' \beta '}\,s_{\beta '}(j)\Big)
-s_\alpha^\dagger(i)\,s_\alpha(i)\,
s_{\alpha '}^\dagger(j)\,s_{\alpha '}(j)\right]\ \ .}
\eqno(2.14)
$$
Due to the constraint (2.11), it is easy to realize
that only the 1 of the first square bracket of (2.14)
yields a non vanishing contribution. In fact the other
three terms, when multiplied by the second square bracket,
give rise to expressions containing annihilation operators
in the same point both for holons and spinons, and hence
vanish when applied to physical states satisfying the
constraint (2.11).
With a similar argument, we can also rewrite the chemical
potential term simply as
$$
\mu~s_\alpha(i)^\dagger\,s_\alpha(i)\ \ .
$$
Thus, the Hamiltonian (2.13) can be drastically simplified
and one gets
$$
\eqalign{
H_{tJ} =& \sum_{<i,j>}
\Bigg\{+\,t~
\Big(e^\dagger(j)\,e(i)\,s_\alpha^\dagger(i)\,
s_\alpha(j)+{\rm h.\,c.}\Big)\cr
&~~~~~~~~~~+{J\over 4}\Big[\sum_{a=1}^3\Big(
s_\alpha^\dagger(i)\,
\sigma^a_{\alpha\beta}\,s_\beta(i)\,
s_{\alpha '}^\dagger(j)\,
\sigma^a_{\alpha ' \beta '}\,s_{\beta '}(j)\Big)
\cr
&~~~~~~~~~~~~~~~~~~~~~-s_\alpha^\dagger(i)\,
s_\alpha(i)\,
s_{\alpha '}^\dagger(j)\,s_{\alpha '}(j)\Big]\Bigg\}
\cr
&+\sum_{i\in \Omega}
\Big[\mu~s_\alpha(i)^\dagger\,s_\alpha(i)+
{\rm i}\,\Lambda(i)\,\phi(i)\Big]\ \ .}
\eqno(2.15)
$$
We stress that the formal manipulations that yielded
the Hamiltonian (2.15) are valid in any dimensions. In the
particular case of two dimensional lattices, it is possible
to obtain (2.15) in an alternative way based on the
abelian bosonization, as shown in [17,18].  The two-dimensional
abelian bosonization consists of replacing the original
fermionic fields by new bosonic degrees of freedom interacting
with an abelian Chern-Simons gauge field [23]. In this
framework, holons and spinons arise, roughly speaking,
as the modulus and phase of the field which bosonizes
the electron operators, and thus at first are both bosonic
\footnote{$^3$}{Actually the correspondence is more involved
than this, and moreover one of the two fields must be a hard-core
boson. For details we refer to the original papers [17,18].}. The
final statistics assignments are dictated by the choice
of the coupling constants between the bosonic holons and
spinons on the one hand, and the abelian Chern-Simons
field on the other.

We now give some details of this construction following [18],
and later we present its  generalization. In the formalism
of second quantization, the gran canonical partition
function for the $t$-$J$ model at temperature $\beta$
and chemical potential $\mu$ is written as an Euclidean
functional integral over a complex Grassmann field
$\psi_\alpha$ with antiperiodic boundary conditions,
according to
$$
{\cal Q}(\beta,\mu) = \int {\cal D}\psi
{}~{\rm e}^{-S(\psi)}
\eqno(2.16)
$$
where
$$
S(\psi) = \int_0^\beta d\tau
\left[\sum_{i\in \Omega} \psi^*_\alpha(i,\tau){\partial
\over {\partial \tau}}
\psi_\alpha(i,\tau)+ {\cal H}_{tJ}(\psi)\right]\ \ .
\eqno(2.17)
$$
In this formula ${\cal H}_{tJ}(\psi)$ denotes the functional
obtained from the Hamiltonian (2.9) by replacing the fermionic
operators
${c'}_\alpha(i)$ and ${c'}_\alpha^\dagger(i)$ with the
Grassmann fields
$\psi_\alpha(i,\tau)$ and $\psi^*_\alpha(i,\tau)$ respectively.
This functional is explicitly given by
$$
\eqalign{
{\cal H}_{tJ}(\psi) =&-t~\sum_{<i,j>}
\Big(\psi_\alpha^*(i,\tau)\,\psi_\alpha(j,\tau)+
{\rm h.\,c.}\Big) + {\cal H}_J(\psi)\cr
&+\mu~\sum_{i\in \Omega}
\psi_\alpha^*(i,\tau)\,\psi_\alpha(i,\tau)
}
\eqno(2.18{\rm a})
$$
where
$$
\eqalign{
{\cal H}_J(\psi)=&{J\over 4}\Bigg[\sum_{a=1}^3
\Big(\psi_\alpha^*(i,\tau)\,
\sigma^a_{\alpha\beta}\,\psi_\beta(i,\tau)\,
\psi_{\alpha '}^*(j,\tau)\,
\sigma^a_{\alpha ' \beta '}\,\psi_{\beta '}(j,\tau)
\Big)\cr
&~~~~~~-\psi_\alpha^*(i,\tau)\,\psi_\alpha(i,\tau)\,
\psi_{\alpha '}^*(j,\tau)\,
\psi_{\alpha '}(j,\tau)\Bigg]\ \ .
}
\eqno(2.18{\rm b})
$$
As is well known, the path integral over anticommuting
Grassmann fields automatically takes into account the
sign factors due the fermionic statistics. An equivalent
expression for ${\cal Q}(\beta,\mu)$ can be obtained by
using, in place of $\psi_\alpha(i,\tau)$ and $\psi_\alpha^*(i,\tau)$,
complex commuting {\it hard core} bosonic fields
$\varphi_\alpha(i,\tau)$ and $\varphi_\alpha^*(i,\tau)$
together with a suitable prescription to implement the
original fermionic statistics. For two dimensional systems,
this prescription is quite simple and elegant: One couples
these bosonic fields to an abelian Chern-Simons gauge field
$A_\lambda$ ($\lambda = 0,1,2$) with level $k=1$, and
averages over all its possible configurations [24].
More precisely, the following identity holds
$$
{\cal Q}(\beta,\mu)= \int {\cal D}\psi
{}~{\rm e}^{-S(\psi)} =\left\langle\int {\cal D}\varphi
{}~{\rm e}^{-{\tilde S}(\varphi; A)}\right\rangle_{k=1}
\eqno(2.19)
$$
where ${\tilde S}(\varphi; A)$ is the bosonized action
obtained from $S(\psi)$
by replacing $\psi_\alpha$ with $\varphi_\alpha$ minimally
coupled to $A_\lambda$. The symbol $\langle~~\rangle_k$
denotes the expectation value over the
Chern-Simons field, namely
$$
\langle\,{\cal O}(A)\,\rangle_k = {{\int {\cal D}A~
{\cal O}(A)~{\rm e}^{-S_{\rm CS}^{k}(A)}}
\over {\int {\cal D}A~{\rm e}^{-S_{\rm CS}^{k}(A)}}}
\eqno(2.20)
$$
where ${\cal O}(A)$ is any functional of $A$ and
$$
S_{\rm CS}^{k}(A)= {k\over{4\pi{\rm i}}}
\int_0^\beta d\tau\int d^2x~
\varepsilon^{\lambda\mu\nu}A_\lambda(\tau,x)
\,\partial_\mu A_\nu(\tau,x)
\eqno(2.21)
$$
is the abelian Chern-Simons action of level $k$.

The bosonization identity (2.19) is essentially
based on the fundamental property that the expectation
values of Wilson loops in Chern-Simons theories are
topological invariants
[24,25]. To see the implications of this fact, let
us consider an arbitrary braid $b$ of $N$ objects
corresponding to the evolution of $N$ particles between
their configurations at Euclidean times $\tau=0$ and
$\tau=\beta$. Due the periodic boundary conditions on $\tau$,
the braid $b$ actually forms a closed link $L(b)$, and
an abelian Wilson loop can be defined as follows
$$
W\Big(L(b),A\Big) = {\rm e}^{\,{\rm i}
\int_{L(b)}\big(A_0(x,\tau)\,d\tau+
\sum\limits_{i=1}^2
A_i(x,\tau)\,dx^i\big)}
\eqno(2.22)
$$
where the line integral in the exponent is
computed alond the link $L(b)$ with a suitable choice of
framing [25]. If $A_\lambda$ is a Chern-Simons
gauge field with action (2.21), one can prove that
$$
\Bigg\langle W\Big(L(b),A\Big)\Bigg\rangle_k =
{\rm e}^{\,{\rm i}\,{{2\pi}\over k}\,n(b)}
\eqno(2.23)
$$
where
$$
n(b) = {1\over 2} \Big[n_+(b)-n_-(b)\Big]
\eqno(2.24)
$$
is the the winding number of the braid $b$,
{\it i.e.} the difference between the number of
overcrossings $n_+(b)$ and the number of
undercrossings $n_-(b)$. If $k=1$, we have
$$
\eqalign{
\Bigg\langle W\Big(L(b),A\Big)\Bigg\rangle_{k=1}
&= {\rm e}^{\,{\rm i}\,\pi\,
\big(n_+(b)-n_-(b)\big)}\cr
&={\rm e}^{\,{\rm i}\,\pi\,
\big(n_+(b)+n_-(b)\big)}\equiv(-1)^{\sigma(\pi(b))}
}
\eqno(2.25)
$$
where $\sigma(\pi(b))$ is the signature of the
permutation $\pi(b)$ associated to the braid $b$.
{}From (2.25) we see that the average over an abelian
Chern-Simons field with level 1 reproduces precisely
the right sign factors needed to implement the
fermionic statistics in terms of purely bosonic variables.
Inserting this result into the gran canonical
partition function and using a path integral
notation, one can finally prove the bosonization
identity (2.19).

When this technique is applied to the $t$-$J$ model,
the electron is described by a hard-core boson
$\varphi_\alpha$ coupled to an abelian Chern-Simons
field. The modulus and phase of $\varphi_\alpha$ are
then interpreted as holon and spinon fields respectively.
More precisely, one writes
$$
\varphi_\alpha(i,\tau) = H^*(i,\tau)\,
\Sigma_\alpha(i,\tau)
\eqno(2.26)
$$
with
$$
\Sigma^*_\alpha(i,\tau)~\Sigma_\alpha(i,\tau)= 1 \ \ .
\eqno(2.27)
$$
Clearly, there is an arbitrariness in the choice
of the phases of $H^*$ and $\Sigma_\alpha$;
in fact if
$$
\eqalign{
H^*(i,\tau)&\rightarrow ~H^*(i,\tau)\,
{\rm e}^{\,-{\rm i} \theta(i,\tau)}\ \ ,\cr
\Sigma_\alpha(i,\tau)&\rightarrow ~
{\rm e}^{\,{\rm i} \theta(i,\tau)}\,
\Sigma_\alpha(i,\tau)\ \ ,}
\eqno(2.28)
$$
the field $\varphi_\alpha$ remains unchanged.
However, modulo this $U(1)$ gauge invariance, one
can still consider $H^*$ and $\Sigma_\alpha$ as the
generalized ``modulus'' and ``phase'' of
$\varphi_\alpha$. Indeed, the constraint (2.27)
expresses the fact that $\Sigma_\alpha$ is confined
on the unit circle like any phase, whereas $H^*$ must
be a hard-core field such that
$$
H^*(i,\tau)~H^*(i,\tau)=0
$$
in order for $\varphi_\alpha$ to be a hard-core
complex bosonic field
as required by the bosonization procedure. $H^*$ and
$\Sigma_\alpha$ are not
yet the most convenient holon and spinon fields, but
these can be obtained with a further change of variables.
In fact, as shown in [18], there are some simplifications
if one introduces new fields ${\tilde E}(i,\tau)$ and
$S_\alpha(i,\tau)$ according to
$$
\eqalign{
H^*(i,\tau)&=\Big(1-{\tilde E}^*(i,\tau)\,
{\tilde E}(i,\tau)\Big)^{-1/2}~{\tilde E}(i,\tau)\ \ ,
\cr
\Sigma_\alpha(i,\tau)&= \Big(1-{\tilde E}^*(i,\tau)\,
{\tilde E}(i,\tau)\Big)^{-1/2}~S_\alpha(i,\tau) \ \ .}
\eqno(2.29)
$$
In terms of these new fields the constraint (2.27)
becomes
$$
{\tilde \Phi}(i,\tau)\equiv {\tilde E}^*(i,\tau)\,
{\tilde E}(i,\tau)+S_\alpha^*(i,\tau)\,
S_\alpha(i,\tau)-1=0
\eqno(2.30)
$$
which is formally similar to (2.11).

In the abelian bosonization of the $t$-$J$ model
proposed in [17,18], the Chern-Simons field $A_\lambda$
is coupled either only to the holons ${\tilde E}$ or
only to the spinons $S_\alpha$. In the first case
the bosonic holons are
transmuted into fermions after integrating out
$A_\lambda$, while the spinons remain bosonic realizing,
in this way, the slave fermion representation.
To see this, let us recall that the bosonized action
of the $t$-$J$ model,
written in terms ${\tilde E}$ and $S_\alpha$, is
$$
\eqalign{
{\tilde S}({\tilde E},S;A)=&
\int_0^\beta d\tau\Bigg\{ \sum_{i\in\Omega}
\Bigg[{\tilde E}^*(i,\tau)\,{\partial\over{\partial \tau}}
{\tilde E}(i,\tau)
+{\rm i} A_0(i,\tau)\Big(1-{\tilde E}^*(i,\tau)\,
{\tilde E}(i,\tau)\Big)\cr
&~~~~~~~~~
+S_\alpha^*(i,\tau)\,{\partial\over{\partial \tau}}
S_\alpha(i,\tau)
+\mu~ S_\alpha^*(i,\tau)\,S_\alpha(i,\tau)
\Bigg]\cr
&-\,t\sum_{<i,j>}
\Bigg[{\tilde E}^*(i,\tau)\,
{\rm e}^{\,{\rm i}\int_{<i,j>}\!\!A_\lambda(x)
\,dx^\lambda}{\tilde E}(j,\tau)
\,S_\alpha^*(i,\tau)\,S_\alpha(j,\tau)
+{\rm h.\,c.}\Bigg]\cr
&~~~~~~~~~~~+{\cal H}_J(S)+\sum_{i\in \Omega}
{\rm i}\,\Lambda(i,\tau)\,
{\tilde \Phi}(i,\tau)\Bigg\}
}
\eqno(2.31)
$$
where $\Lambda(i,\tau)$ is the Lagrange multiplier
enforcing the constraint (2.30), and
${\cal H}_J(S)$ is given by (2.18b) with
$\psi_\alpha(i,\tau)$ and $\psi_\alpha^*(i,\tau)$ replaced by
$S_\alpha(i,\tau)$ and $S_\alpha^*(i,\tau)$
respectively.
As mentioned before, in (2.31) the Chern-Simons
field is coupled (in the standard minimal
way) only to the holon fields ${\tilde E}$ and
${\tilde E}^*$. Using this action and (2.19),
the partition function of the $t$-$J$ model is
formally given by
$$
{\cal Q}(\beta,\mu) = \left\langle \int{\cal D}{\tilde E}
\,{\cal D}S ~
{\rm e}^{-{\tilde S}({\tilde E},S;A)}\right\rangle_{k=1}\ \ .
\eqno(2.32)
$$
Finally, after computing the expectation value
over the Chern-Simons field (or equivalently after
eliminating $A_\lambda$
through its field equations) one gets the effective
action of $t$-$J$ model in the slave fermion
representation, {\it i.e.}
$$
\eqalign{
S_{\rm eff}(E,S)=&
\int_0^\beta d\tau\Bigg\{ \sum_{i\in\Omega}\Bigg[E^*(i,\tau)\,
{\partial\over{\partial \tau}}E(i,\tau)
+S_\alpha^*(i,\tau)\,{\partial\over{\partial \tau}}
S_\alpha(i,\tau)\Bigg]\cr
&~~~~~~-\,t\sum_{<i,j>}
\Big[E^*(i,\tau)\,
E(j,\tau)
\,S_\alpha^*(i,\tau)\,S_\alpha(j,\tau)
+{\rm h.\,c.}\Big]\cr
&~~~~~+{\cal H}_J(S)+\sum_{i\in \Omega}\Big[\mu~
S_\alpha^*(i,\tau)\,S_\alpha(i,\tau)
+{\rm i}\,\Lambda(i,\tau)\,
\Phi(i,\tau)\Big]\Bigg\}\ \ .
}
\eqno(2.33)
$$
Here $E$ and $E^*$ denote the hard-core holon fields
dressed by the effective Chern-Simons contribution.
They are defined as follows
$$
E(i,\tau) \equiv ~{\rm e}^{\,{\rm i}\sum\limits_{j\in \Omega}
\Theta(i,j)\,\big(
{\tilde E}^*(j,\tau)\,{\tilde E}(j,\tau) - 1\big)}~
{\tilde E}(i,\tau)
\eqno(2.34)
$$
where $\Theta(i,j)$ is the lattice angle function
(see [26,27,28] and Section 4 for a detailed discussion
of its properties). Using the canonical {\it commutation}
relations of ${\tilde E}$ and ${\tilde E}^*$, it
is easy to show
that $E$ and $E^*$ are fermions satisfying the
equal time {\it anticommutation} relations
$$
\Big\{ E(i,\tau)\,,\,E^*(j,\tau)\Big\} = \delta(i,j)\ \ .
\eqno(2.35)
$$
Furthermore, from (2.34) it is clear that
$$
E^*(i,\tau)\,E(i,\tau)={\tilde E}^*(i,\tau)\,
{\tilde E}(i,\tau)\ \ ,
$$
and hence
$$
\Phi(i,\tau)\equiv  E^*(i,\tau)\,E(i,\tau)+S_\alpha^*(i,\tau)
\,S_\alpha(i,\tau) -1 ={\tilde \Phi}(i,\tau)\ \ .
$$

Using these results, it becomes evident that the gran
canonical partition function can be written as a path
integral over both fermionic and bosonic fields according to
$$
{\cal Q}(\beta,\mu) = \int{\cal D}E\,{\cal D}S ~
{\rm e}^{-S_{\rm eff}(E,S)}\ \ .
\eqno(2.36)
$$
We observe that
this path integral implements correctly the
fermionic statistics of the original electronic problem.
{}From the effective action
(2.33) one can easily derive the Hamiltonian of the $t$-$J$ model
in the slave fermion description; when written in the
operator formalism, such Hamiltonian is exactly the same
as the one given in (2.15), which we obtained by simply
inserting (2.10) into (2.9).

As remarked in [18], also the slave boson description of
the $t$-$J$ model can be realized via abelian bosonization.
In fact, if we start from an action similar to (2.31)
but with the Chern-Simons field $A_\lambda$ coupled
only to the spinons $S_\alpha$, upon elimination of
$A_\lambda$, these are transmuted into fermions while
the holons remain bosons.
Using the bosonization formalism, it becomes apparent
that the slave fermion and the slave boson descriptions
are deeply related to each other and can be viewed as
two different gauge fixings of an abelian gauge
invariance exhibited by the bosonized Hamiltonian
of the $t$-$J$ model.

The authors of [18] considered also a non abelian
variant of this bosonization technique by introducing
two Chern-Simons gauge fields: One abelian, and one
in the fundamental representation of $SU(2)$. In order
to represent correctly the fermionic statistics of the
electronic problem, the coefficients of the two
Chern-Simons
terms must be judiciously chosen (see {\it e.g.}
Eq. (3.10) of [18]). Furthermore, in order
to describe only abelian representations of the
braid group ({\it i.e.} fields whose statistics
is simply described by a phase and not by a braiding
matrix), the level $k$ of the $SU(2)$ Chern-Simons
term must be related to the spin $s$ of the $SU(2)$
representation according to
$k=\pm2\,s$.
As mentioned, the $SU(2)$ field is taken in the
fundamental representation with $s=1/2$ (the
non-relativistic electrons are indeed a doublet
of $SU(2)$), and thus $k$ must be either $+1$
or $-1$.
In this non abelian bosonization, the abelian
field is coupled to the (bosonic) holons while
the $SU(2)$ field is coupled to the (bosonic)
spinons; integrating out these Chern-Simons fields,
holons and spinons are transmuted into semions.

Now we are going to generalize (and to simplify)
the construction of [18] by showing that the semionic
statistics of holons and spinons can be realized
within the context of the abelian bosonization without
using non abelian Chern-Simons fields. More generally,
we will show that holons and spinons in the two-dimensional
$t$-$J$ model can be arbitrary anyons subject only to
the constraint that the composite object made of one
holon and one spinon be a fermion. We call this the
slave anyon representation of the $t$-$J$ model which,
as we will show later, smoothly interpolates between
the slave fermion and the slave boson
descriptions that we mentioned earlier.
Once again we stress that this ``anyonization'' is
a purely formal operation based on
mathematical manipulations on the partition function;
to see which statistics actually occurs, requires a
serious investigation of the dynamical aspects of
the model.

Let us now give some details of our construction.
We start from the effective action (2.33) in the
slave fermion picture and, as in [18], we introduce
two independent Chern-Simons fields, $B_\lambda$
and $V_\lambda$, but, contrary to [18], we
take both of them abelian \footnote{$^4$}{One
could equivalently consider the action (2.31)
and introduce only a new Chern-Simons field
$V_\lambda$ coupled to $S_\alpha$ and suitably
change the coupling constant in $S_{\rm CS}(A)$.}.
Then we minimally couple
the fermionic holon field $E$ only to $B_\lambda$
and the bosonic spinon field $S_\alpha$ only to
$V_\lambda$. The resulting action is
$$
\eqalign{
{\tilde S}(E,S;B,V) &=
\int_0^\beta d\tau\Bigg\{ \sum_{i\in\Omega}
\Bigg[ E^*(i,\tau)\,{\partial\over{\partial \tau}}
E(i,\tau) + {\rm i}\, B_0(i,\tau)
\Big(1-E^*(i,\tau)\,E(i,\tau)\Big)
\cr
&~~~~~~~~~~~~~~~~~~+S_\alpha^*(i,\tau)\,
{\partial\over{\partial \tau}}
S_\alpha(i,\tau)+{\rm i}\, V_0(i,\tau)
\,S_\alpha^*(i,\tau)\,S_\alpha(i,\tau)
\Bigg]\cr
&~~~~~~~~~~~~~-\,t\sum_{<i,j>}
\Bigg[E^*(i,\tau)\,
{\rm e}^{\,{\rm i}\int_{<i,j>}\!\!
B_\lambda(x)\,dx^\lambda}E(j,\tau)~\times\cr
&\hskip 3.5cm \times ~
S_\alpha^*(i,\tau)\,{\rm e}^{\,{\rm i}
\int_{<i,j>}\!\!{V}_\lambda(x)\,
dx^\lambda}S_\alpha(j,\tau)
+{\rm h.\,c.}\Bigg]\cr
&~~~~~~~~~~~+{\cal H}_J(S)+\sum_{i\in \Omega}
\Big[\mu~S_\alpha^*(i,\tau)\,S_\alpha(i,\tau)+
{\rm i}\,\Lambda(i,\tau)\,
\Phi(i,\tau)\Big]\Bigg\}\ \ .
}
\eqno(2.37)
$$
Notice that the time component $B_0$ is
coupled to $(1-E^*\,E)$ and not simply to the
density $E^*\,E$: This is not an artifact but is
a direct consequence of expressing the original
action of the $t$-$J$ model in terms of $E$ and
$S_\alpha$ (cf Eq. (2.31)). The other couplings
with the gauge fields are no surprise, and one
easily sees that
$$
{\tilde S}(E,S;B,V)\Big|_{B=V=0} =
S_{\rm eff}(E,S)\ \ .
$$
In the spirit of the bosonization procedure,
there should be no difference even when the two
Chern-Simons fields are non vanishing. In particular,
the introduction of $B_\lambda$ and $V_\lambda$,
with their corresponding actions at level $k_B$
and $k_V$ respectively (see Eq. (2.21)), should
not spoil the overall fermionic statistics that
is already implemented in a correct way by the
effective action (2.33). Therefore the levels of
the two Chern-Simons terms must be carefully chosen.
Using (2.23), it is not difficult to realize that
the correct statistics is mantained if
$$
k_B=-k_V ~\,{\rm mod}\,2\ \ .
\eqno(2.38)
$$
The minimal choice is
$$
k_B=-k_V = {1\over \nu}
\eqno(2.39)
$$
where the parameter $\nu$ is introduced for later
convenience. Contrarily to the non abelian bosonization
of [18] where the levels of the Chern-Simons terms
are both
fixed, in our approach there is a free parameter which,
as we will see in a moment, allows to realize anyons of
arbitrary statistics and not only semions.

In this approach the gran canonical partition
function is
$$
{\cal Q}(\beta,\mu) = \Big\langle\Big\langle
\int {\cal D}E \,{\cal D}S
{}~{\rm e}^{-{\tilde S}(E,S;B,V)}\Big\rangle
\Big\rangle_\nu
\eqno(2.40)
$$
where the symbol $\langle\langle~~~\rangle\rangle_\nu$
denotes the expectation value with respect to the
Chern-Simons fields $B_\lambda$ and $V_\lambda$
whose levels are given as in (2.39). Eliminating
$B_\lambda$ and $V_\lambda$ through their field
equations, the resulting action can be entirely
expressed in terms
of new fields, ${\hat E}$ and ${\hat S}_\alpha$ and
their complex conjugates, defined as follows
$$
\eqalign{
{\hat E}(i,\tau) &\equiv ~{\rm e}^{\,{\rm i}\,\nu\,
\sum\limits_{j\in \Omega}\Theta(i,j)\,\big(
E^*(j,\tau)\,E(j,\tau) - 1\big)}~E(i,\tau)\ \ ,\cr
{\hat S}_\alpha(i,\tau) &\equiv ~{\rm e}^{\,-{\rm i}
\,\nu\,\sum\limits_{j\in \Omega}\Theta(i,j)\,
S_\alpha^*(j,\tau)\,S_\alpha(j,\tau) }~S_\alpha(i,\tau)\ \ .
}
\eqno(2.41)
$$
These are anyonic fields of statistics $\nu$ and $-\nu$
respectively, and, as such, satisfy braiding relations
among themselves. We postpone the detailed discussion of
their properties to Section 4; here we simply notice that
${\hat E}$ is a fermion based anyon
whilst ${\hat S}_\alpha$ is boson based. The effective
action resulting from (2.40) realizes the slave anyon
representation of the $t$-$J$ model and is essentially
given by (2.33) with $E$ and $S_\alpha$
replaced by ${\hat E}$ and ${\hat S}_\alpha$.
However, this representation is purely formal
because the anyonic fields (2.41) are
{\it non local}.
In fact, the angle function appearing in the
exponents of (2.41) is multivalued and, to remove
all ambiguities, a cut and a fundamental domain must
be chosen implying that the resulting fields
become non-local [29].
Of course, if $\nu=1$ or $\nu=0$ these problems
cease to exist since in these circumstances the
anyons becomes ordinary bosons or fermions. These two
cases are clearly in direct correspondence with the
slave fermion and slave boson representations that we
discussed earlier, and thus our slave anyon representation
smoothly interpolates between them.
It is worth to point out that the effective action
$S_{\rm eff}({\hat E},{\hat S}_\alpha)$
is well-defined and unambiguous for any value of
$\nu$ and does not depend on the detailed definition
of the angle function; however due to the non
locality
of ${\hat E}$ and ${\hat S}_\alpha$, it is
problematic to path-integrate over them.
Therefore, in the functional formalism the slave
anyon representation must be written in terms of
bosonic and fermionic fields coupled to two abelian
Chern-Simons fields as in (2.37), with the gran
canonical partition function given by (2.40).
Things are different in the operator formalism
where one can define generalized commutation
relations for anyonic operators. These are actually
braiding relations which depend crucially on how the
angle function is defined, as we will discuss in
Section 4.

\vfill \eject
\centerline{\bf 3. The Supersymmetric $t$-$J$
Model}
\vskip 0.7cm
When the coupling constants and the chemical potential
of the $t$-$J$ model are
$$
J=2\,t~~~~,~~~~\mu=\gamma(d)\,t
\eqno(3.1)
$$
where $\gamma(d)$ is the coordination number
of the $d$-dimensional lattice $\Omega$,
\footnote{$^5$}{Given any lattice site $i\in
\Omega$, the coordination number is defined as the
number of points which are nearest neighbors to $i$.
For example, for a square lattice one has $\gamma(d)=2\,d$.}
the Hamiltonian (2.1) enjoys very special properties: It
can be written as a graded permutation operator (up to
an irrelevant constant term proportional to the number
of lattice points) [30] and it commutes with the
generators of a superalgebra $SU(1|2)$ [6]. In
$d=1$ the $t$-$J$ model at the supersymmetric
point (3.1) can be exactly solved by Bethe {\it
Ansatz} [7,8], and it is tempting to speculate
that also in higher dimensions the supersymmetry
has some non trivial implications.

Let us first observe that there is a very simple
realization of the superalgebra $SU(1|2)$ in terms
of the projected fermionic operators (2.6)
introduced in Section 2.
In fact, defining
$$
\eqalign{
J^+(i) &= {c'}^\dagger_+(i)\,{c'}_-(i)
{}~~~~,~~~~
J^-(i) = {c'}^\dagger_-(i)\,{c'}_+(i)\ \ ,
\cr
J^0(i) &= {1\over 2}\Big({n'}_+(i)-{n'}_-(i)
\Big)  \ \ ,\cr
T(i) &= 1-{1\over 2}\Big({n'}_+(i)+{n'}_-(i)
\Big)\ \ ,\cr
Q_\alpha(i) &= {c'}_\alpha(i)~~~~,~~~~
Q^\dagger_\alpha(i) = {c'}^\dagger_\alpha(i)
{}~~~~~~(\alpha=\pm 1)\ \ ,
}\eqno(3.2)
$$
and using (2.7), it is immediate to verify that
$$
\eqalign{
\Big[ J^+\,,\,J^-\Big] &= 2\,J^0 ~~~~,~~~~
\Big[ J^0\,,\,J^\pm\Big] = \pm\,J^\pm \ \ ,\cr
\Big\{ Q_\alpha^\dagger\,,\,Q_{-\alpha}\Big\}
&= \,J^\alpha~~~~,~~~~
\Big\{ Q_\alpha^\dagger\,,\,Q_\alpha\Big\} =
\alpha\,J^0+T\ \ ,\cr
\Big[ J^0\,,\,Q_\alpha\Big] &=-{\alpha\over 2}
\,Q_\alpha ~~~~,~~~~
\Big[ J^0\,,\,Q_\alpha^\dagger\Big] ={\alpha\over 2}
\,Q_\alpha^\dagger\ \ ,\cr
\Big[ T\,,\,Q_\alpha\Big] &={1\over 2}\,Q_\alpha
{}~~~~,~~~~
\Big[ T\,,\,Q_\alpha^\dagger\Big] =-{1\over 2}\,
Q_\alpha^\dagger \ \ ,
}
\eqno(3.3)
$$
where $J^+=\sum\limits_{i\in\Omega}J^+(i)$, and so
on. These are precisely the graded commutation relations
of the superalgebra $SU(1|2)$, which in some circles is
also denoted by $spl(2,1)$. Notice that $Q_\alpha$ and
$Q_\alpha^\dagger$ for $\alpha=\pm1$ form {\it two}
doublets of fermionic generators; and hence one can
say that (3.3) is a $N=2$ supersymmetry algebra [31].

Choosing the coupling constants of $H_{tJ}$ as in
(3.1), it is easy to check that
$$
\Big[T^A\,,\,H_{tJ}\Big] = 0
\eqno(3.4)
$$
where $T^A$ is any one of the eight generators of
$SU(1|2)$. Thus the $t$-$J$ model with
$J=2\,t$ and $\mu=\gamma(d)\,t$ is manifestly $N=2$
supersymmetric in any dimension $d$.
We denote the corresponding Hamiltonian by
$H_{tJ}^{\rm SUSY}$. It is interesting to observe
that the supersymmetry is non-linearly realized
in $H_{tJ}^{\rm SUSY}$; in fact the action of the
generators $T^A$ on the projected electronic
operators yields expressions which, in general,
are non-linear. For example one finds
$$
\Big\{ Q_+\,,\,{c'}_-^\dagger(i)\Big\}=
{c'}_-^\dagger(i)\,{c'}_+(i)\ \ .
\eqno(3.5)
$$
We now show that there is a simpler way to describe
this supersymmetry by using the slave fermion or the
slave boson representations with which it can be
realized in a {\it linear} way. To avoid repetitions,
here we limit our discussion to the slave fermion case,
while in the next section we will discuss in detail the
supersymmetry in the slave anyon representation.

Let us then consider the fermionic holon operators
$e(i)$ and $e^\dagger(i)$, and the bosonic spinon
operators $s_\alpha(i)$ and $s^\dagger_\alpha(i)$
which satisfy the (anti)commutation relations (2.12),
and let us define
$$
\eqalign{
J^+&=\sum_{i\in \Omega}s^\dagger_+(i)\,s_-(i)~~~~,~~~~
J^-=\sum_{i\in \Omega}s^\dagger_-(i)\,s_+(i) \ \ ,\cr
J^0&=\sum_{i\in \Omega}{1\over 2}\Big(s^\dagger_+(i)\,s_+(i)-
s^\dagger_-(i)\,s_-(i)\Big)
\ \ , }
\eqno(3.6{\rm a})
$$
$$
T=\sum_{i\in \Omega}\Big(e^\dagger(i)\,
e(i)+{1\over 2}\,s^\dagger_\alpha(i)\,
s_\alpha(i)
\Big)\ \ ,
\eqno(3.6{\rm b})
$$
$$
Q_\alpha=\sum_{i\in \Omega}e^\dagger(i)\,
s_\alpha(i)~~~~,~~~~
Q^\dagger_\alpha=\sum_{i\in \Omega}e(i)\,
s^\dagger_\alpha(i)~~~~~(\alpha=\pm 1)\ \ .
\eqno(3.6{\rm c})
$$
Using (2.12), it is very easy to check that these
eight operators close the superalgebra
$SU(1|2)$ given in (3.3). Furthermore, the action
of these generators on holons and
spinons is always linear. In particular, the $SU(2)$
generators $J^\pm$ and $J^0$ commute with $e(i)$ and
$e^\dagger(i)$, which are indeed scalars under
rotations, while on $s_\alpha(i)$ and
$s_\alpha^\dagger(i)$ they act as follows
$$
\eqalign{
\Big[J^0\,,\,s^\dagger_\alpha(i)\Big]&={\alpha\over 2}
\,s^\dagger_\alpha(i)
{}~~~~,~~~~\Big[J^0\,,\,s_\alpha(i)\Big]=-{\alpha\over 2}
\,s_\alpha(i)\ \ ,\cr
\Big[J^\alpha\,,\,s^\dagger_{-\alpha}(i)\Big]&=
s^\dagger_\alpha(i)~~~~,~~~~
\Big[J^\alpha\,,\,s_\alpha(i)\Big]=-s_{-\alpha}(i)\ \ .
}
\eqno(3.7{\rm a})
$$
(Of course, in the last two commutators there is no
sum over $\alpha$.)
These commutation relations simply express the fact that
$s_\alpha(i)$ and $s_\alpha^\dagger(i)$ for $\alpha=\pm 1$
are two spin 1/2 doublets of $SU(2)$. The abelian
generator $T$ measures the $U(1)$ charge of holons and
spinons according to
$$
\eqalign{
\Big[T\,,\,s_\alpha(i)\Big]&=-{1\over 2}s_\alpha(i)
{}~~~~,~~~~
\Big[T\,,\,s^\dagger_\alpha(i)\Big]={1\over 2}
s^\dagger_\alpha(i)\ \ ,\cr
\Big[T\,,\,e(i)\Big]&=-e(i)~~~~,~~~~
\Big[T\,,\,e^\dagger(i)\Big]=e^\dagger(i)\ \ .
}
\eqno(3.7{\rm b})
$$
Finally, the supersymmetry generators $Q_\alpha$
and $Q_\alpha^\dagger$ transform the fermionic holons
into the bosonic spinons and vice-versa. More precisely
one finds
$$
\eqalign{
\Big[Q_\alpha\,,\,s^\dagger_\beta(i)\Big]
&= e^\dagger(i)\,\delta_{\alpha\beta}
{}~~~~,~~~~
\Big[Q^\dagger_\alpha\,,\,s_\beta(i)\Big] =
-e(i)\,\delta_{\alpha\beta}\ \ ,\cr
\Big\{Q_\alpha\,,\,e(i)\Big\}&=s_\alpha(i)~~~~,
{}~~~~
\Big\{Q^\dagger_\alpha\,,\,e^\dagger(i)\Big\} =
s^\dagger_\alpha(i)\ \ ,
}
\eqno(3.7{\rm c})
$$
with the other possible (anti)commutators vanishing.

Before looking at the invariance properties of the
Hamiltonian, let us observe that
the constraint $\phi(i)$ in (2.11) is invariant under
$SU(1|2)$, {\it i.e.}
$$
\Big[T^A\,,\,\phi(i)\Big] = 0\ \ .
\eqno(3.8)
$$
This property is certainly welcome and will have very
important implications in the following.
Let us now turn to the Hamiltonian $H_{tJ}$ in the slave
fermion representation
given by (2.15), and chose the coupling constants as in
(3.1). A simple calculation shows that this Hamiltonian
commutes with the generators of $SU(1|2)$ only if the
Lagrange multiplier $\Lambda(i)$ transforms in a non
trivial way under the supersymmetries.
To be precise one finds that
$$
\Big[Q_\alpha\,,\,H_{tJ} \Big]=0
$$
only if
$$
\Big[Q_\alpha\,,\,\Lambda(i)\Big] =
-{\rm i}\,{J\over 2}\,\Big(\Delta+\gamma(d)\Big)\,
e^\dagger(i)\,s_\alpha(i)
\eqno(3.9)
$$
where $\Delta$ is the lattice Laplace operator. Even
though acceptable on general grounds, this transformation property is rather
complicated, especially
when compared with the ones of the other operators
entering the Hamiltonian (see (3.7)).
Moreover it poses serious problems in setting up a mean
field theory without breaking
the supersymmetry. Indeed, the mean field approximation
(MFA) instructs us to replace the operator $\Lambda(i)$
with its mean value $\lambda(i)$ which is a {\it number}
and, as such, commutes with the supersymmetry generators.
Therefore, if $\Lambda(i)$ has non trivial transformation properties like
(3.9), the MFA certainly breaks the supersymmetry. This is
very unsatisfactory because the
mean field approximation is one of the most viable methods
to analyze the theory
in dimensions bigger than one, and if (3.9) were unavoidable,
it would seem that the role of the supersymmetry of the $t$-$J$
model can not be investigated in this approach.

However, this problem can be overcome. Indeed, since the
Hilbert space of the $t$-$J$ model is characterized by
$\phi(i)=0$, we can add to the Hamiltonian (2.15) any
expression
proportional to $\phi(i)$ without changing the physical
dynamics of the system. Thus
we can try to modify the Hamiltonian $H_{tJ}$ in such
a way that the supersymmetry
be realized without transforming the Lagrange multiplier.
This is possible if we consider the following operator
$$
H'_{tJ} = H_{tJ} +{J\over 4} \sum_{<i,j>}
\Bigg[\Big(s^\dagger_\alpha(i)\,
s_\alpha(i)
-e^\dagger(i)\,e(i)\Big)\,\phi(j) +
\Big(i\leftrightarrow j\Big) \Bigg]
\eqno(3.10)
$$
where $H_{tJ}$ is the Hamiltonian (2.15) with the
coupling constants chosen according to (3.1). Because
of our previous observation, $H'_{tJ}$ is equivalent to
$H_{tJ}$, since their difference vanishes on the
physical states.
With a straightforward calculation, one can easily
prove that the new Hamiltonian $H'_{tJ}$ commutes
with all generators of $SU(1|2)$ without requiring
any transformation property of the Lagrange multiplier
$\Lambda(i)$, namely
$$
\Big[T^A\,,\,H'_{tj}\Big]=0 ~~~~{\rm and}~~~~
\Big[T^A\,,\,\Lambda(i)\Big]=0\ \ .
\eqno(3.11)
$$
Now the mean field approximation can be used on
$H'_{tJ}$ {\it without} breaking the
supersymmetry. Actually, we can go even further.
With simple algebra, the Hamiltonian
(3.10) can be rewritten as follows
\footnote{$^6$}{Here and in the following,
for any operator ${\cal O}$, the symbol
$\big|{\cal O}\big|^2$ means ${\cal O}^\dagger\,{\cal O}$.}
$$
\eqalign{
H'_{tJ} =& {J\over 2}\sum_{<i,j>}
\Big|s^\dagger_\alpha(i)\,s_\alpha(j)+
e^\dagger(i)\,e(j)\Big|^2 \cr
&+ \sum_{i\in \Omega}\Bigg[{J\over 4}\,\gamma(d)\Big(
s^\dagger_\alpha(i)\,s_\alpha(i)+e^\dagger(i)\,e(i)\Big)
+{\rm i}\,\Lambda(i)\,\phi(i)\Bigg] \ \ .}
\eqno(3.12)
$$
Using (2.11), we then have
$$
H'_{tJ} = {J\over 2}\sum_{<i,j>}
\Big|s^\dagger_\alpha(i)\,s_\alpha(j)+
e^\dagger(i)\,e(j)\Big|^2
+ \sum_{i\in \Omega}{\rm i}\,\Lambda'(i)\,
\phi(i) -{J\over 4}\,\gamma(d)\,V
\eqno(3.13)
$$
where
$$
\Lambda'(i)=\Lambda(i) -{\rm i}\, {J\over 4}\,
\gamma(d)\ \ ,
$$
and $V$ is the volume of $\Omega$, {\it i.e.} the
number of lattice points.

This last form of the Hamiltonian is particularly
simple and suggestive. In fact, using (3.7) it is easy
to see that the expression inside the absolute value sign,
$$
G(i,j)=s^\dagger_\alpha(i)\,s_\alpha(j)+e^\dagger(i)\,e(j)
\eqno(3.14)
$$
is invariant under $SU(1|2)$. Then, since the constraint
$\phi(i)$ is also invariant, $H'_{tJ}$ commutes with
the generators of $SU(1|2)$ only if the Lagrange
multiplier $\Lambda'(i)$ does not transform, as desired.
Moreover, when written as
in (3.13), the Hamiltonian of the $t$-$J$ model appears
as the natural generalization of the Heisenberg XXX model.
Indeed, if we delete all $e$ and $e^\dagger$ operators
from (3.13), we recover precisely the antiferromagnetic
Heisenberg model which, as well
known, is invariant under $SU(2)$. Thus, we could say
that the supersymmetric $t$-$J$ model is the natural
generalization of the Heisenberg model, when the symmetry
algebra $SU(2)$ is generalized to $SU(1|2)$. A further
generalization of the $t$-$J$ model in this respect, is
given by the model recently proposed in [31], which is
invariant under
the superalgebra $SU(2|2)$. In fact, when written in the
slave fermion representation,
the Hamiltonian of [31] is like (3.13) but with two sets
of fermionic oscillators, corresponding respectively to
holons and localons. The latter are possible because in
this model the double electronic occupancy on the same site
is allowed and local pairs of electrons (localons) can arise
in the spectrum.

The supersymmetric Hamiltonian (3.13) is particularly
suited for a further simplification: We can perform
on it the Hubbard-Stratonovich transformation [32]
to simplify the $J$-term, which is quartic in the
operators, without breaking the supersymmetry. In
fact, let us introduce a complex bosonic
Hubbard-Stratonovich operator
$F_{ij}$ defined on the link between the nearest neighbor
points $i$ and $j$, and consider the Hamiltonian
$$
\eqalign{
{\tilde H}'_{tJ} = &\sum_{<i,j>}\Bigg[
-{2\over J}\,F_{ij}^\dagger\,F_{ij}+
F^\dagger_{ij}\,\Big(
s^\dagger_\alpha(i)\,s_\alpha(j)+e^\dagger(i)
\,e(j)\Big)\cr
&~~~~~~~~~+\Big(s^\dagger_\alpha(j)\,s_\alpha(i)
+e^\dagger(j)\,e(i)
\Big)\,F_{ij}\Bigg]
+\sum_{i\in\Omega}{\rm i}\,\Lambda'(i)\,\phi(i) -
{J\over 4}\,\gamma(d)\,V\ \ .
}
\eqno(3.15)
$$
Upon elimination of $F_{ij}$ through its ``equation of
motion'', ${\tilde H}'_{tJ}$ becomes equivalent to
$H'_{tJ}$ and thus it describes the same dynamics.
The advantage of writing the Hamiltonian as in (3.15)
is that the coupling among the operators is much simpler
than in (3.13): There are no more quartic interactions.
Moreover, since the quantity in (3.14) is invariant
under $SU(1|2)$ as we have previously remarked,
${\tilde H}'_{tJ}$ is invariant only if the
Hubbard-Stratonovich operator does not transform,
{\it i.e.}
$$
\Big[T^A\,,\,F_{ij}\Big]=0\ \ .
\eqno(3.16)
$$
Again this property is welcome, because it allows
to make the MFA on ${\tilde H}'_{tJ}$ without
breaking supersymmetry. In fact, we can replace
the operators $F_{ij}$ and $\Lambda'(i)$ with
their corresponding mean values $f_{ij}$ and
$\lambda'(i)$, determined in a self-consistent
way, and get the mean field Hamiltonian
$$
\eqalign{
{\tilde H}_{\rm MF} = &\sum_{<i,j>}\Bigg[
-{2\over J}\,f_{ij}^*\,f_{ij}+f^*_{ij}\,\Big(
s^\dagger_\alpha(i)\,s_\alpha(j)+
e^\dagger(i)\,e(j)\Big)\cr
&~~~~~~~~~+\Big(s^\dagger_\alpha(j)\,s_\alpha(i)
+e^\dagger(j)\,e(i)
\Big)\,f_{ij}\Bigg]
+\sum_{i\in\Omega}{\rm i}\,\lambda'(i)\,\phi(i) -
{J\over 4}\,\gamma(d)\,V
}
\eqno(3.17)
$$
which is invariant under $SU(1|2)$.

A few remarks are in order. First of all, once the
MFA is made and in particular $\Lambda'(i)$ is
replaced by $\lambda'(i)$, the constraint
$\phi(i)=0$ is not enforced any more. Then, the
condition that only the states without double
electronic occupancy are permitted is released,
and in principle new states and excitations, where
the number of holons and spinons is not restricted
as in (2.11), become allowed. Secondly, one can
improve the MFA without breaking supersymmetry if
one considers the phase of the complex field $f_{ij}$
and the amplitude of $\lambda'(i)$ as fluctuating.
In other words, one writes
$$
\eqalign{
f_{ij} &= {\tilde f}_{ij}~
{\rm e}^{{\rm i}\,A_{ij}}\ \ ,\cr
\lambda'(i)&={\tilde \lambda}(i)+A(i)}
\eqno(3.18)
$$
where ${\tilde f}_{ij}$ and ${\tilde \lambda}(i)$
are fixed complex numbers while $A_{ij}$ and $A(i)$
are dynamical variables. In particular, if one chooses
${\tilde f}_{ij}={\tilde f}$ and $\lambda(i)={\tilde \lambda}$,
where ${\tilde f}$ and ${\tilde \lambda}$ are real constants,
one selects the so-called {\it uniform phase} of
the slave fermion representation [13,20], whose
Hamiltonian is \footnote{$^7$}{We drop from ${\tilde H}$
an irrelevant additive constant.}
$$
\eqalign{
{\tilde H} =&~{\tilde f}~\sum_{<i,j>}
\Bigg(
s^\dagger_\alpha(i)\,{\rm e}^{{\rm i}\,A_{ij}}
\,s_\alpha(j)+e^\dagger(i)\,{\rm e}^{{\rm i}
\,A_{ij}}\,e(j)+{\rm h.c.}\Bigg)\cr
&+\sum_{i\in\Omega}{\rm i}\,A(i)~
\Big(s^\dagger_\alpha(i)\,s_\alpha(i) +
e^\dagger(i)\,e(i)\Big)\cr
&+{\rm i}\,{\tilde \lambda}~\sum_{i\in\Omega}
\Big(s^\dagger_\alpha(i)\,s_\alpha(i) +
 e^\dagger(i)\,e(i)\Big)\ \ .
}
\eqno(3.19)
$$
The first two lines can be interpreted as
the Hamiltonian of a non relativistic fermion
(holon) and a pair of non relativistic spin 1/2
bosons (spinons) minimally coupled to an Abelian gauge
field whose components $A_\mu(i)$ are
$$
A_0(i) = A(i)~~~~,~~~~A_1(i)=A_{i,i+{\hat 1}}~~~~,~~~~
A_2(i)=A_{i,i+{\hat 2}}
\eqno(3.20)
$$
where ${\hat 1}$ and ${\hat 2}$ denote the unit lattice
vectors along the $x$- and $y$-axis respectively. This gauge
field can be identified with the resonating
valence bond (RVB) field introduced in [2,3,11]. A system
similar to that described by (3.19) was also considered and
analyzed in [33], but there the masses of the fermions and
bosons were taken to be different, thus breaking explicitly
the supersymmetry.

We conclude this section with a few comments. First of all,
the discussion we have presented using the operators in the
slave fermion representation can be easily repeated in the
slave boson formalism, where the holons are bosonic and the
spinons fermionic. In this case most of the formulas, in
particular the Hamiltonians (3.13), (3.15), (3.17) and (3.19),
remain unchanged. The only formal difference is in the action
of the supersymmetry generators on holons and spinons, since,
changing the statistics, the commutators must be replaced by
anticommutators and vice-versa. Indeed, Eqs. (3.7c) are
replaced by
$$
\eqalign{
\Big\{Q_\alpha\,,\,s^\dagger_\beta(i)\Big\} &=
e^\dagger(i)\,\delta_{\alpha\beta}
{}~~~~,~~~~
\Big\{Q^\dagger_\alpha\,,\,s_\beta(i)\Big\} =
e(i)\,\delta_{\alpha\beta}\ \ ,\cr
\Big[Q_\alpha\,,\,e(i)\Big]&=-s_\alpha(i)~~~~,~~~~
\Big[Q^\dagger_\alpha\,,\,e^\dagger(i)\Big] =
s^\dagger_\alpha(i)\ \ .
}
\eqno(3.21)
$$
On the contrary, (3.7a-b) remain valid also in the
slave boson formalism, because $J^\pm$, $J^0$ and $T$
are bosonic generators which always close commutators
independently of the statistics of the operators on which they act.

Finally, we observe that the analysis of the
supersymmetry properties of the $t$-$J$ model as
presented in this section, can be equivalently formulated
in the functional approach, using fields instead of operators.
In this case, the action of the generators of the superalgebra
$SU(1|2)$ on holons and spinons is expressed as transformation
laws of the corresponding fields, instead of (anti)commutators.
For example, using the notations of Section 2, the infinitesimal
variations of the holon and spinon fields turn out to be
$$
\eqalign{
\delta E(i,\tau) &= -{{}\over{}}\eta\,E(i,\tau) +
\epsilon_+
\,S_+(i,\tau)+\epsilon_-
\,S_-(i,\tau)\ \ , \cr
\delta E^*(i,\tau) &= +{{}\over{}}\eta\,E^*(i,\tau) +
\epsilon^*_+
\,S^*_+(i,\tau)+\epsilon^*_-
\,S^*_-(i,\tau)\ \ , \cr
\delta S_+(i,\tau) &=-{\theta^0\over 2}\,S_+(i,\tau)
-\theta^+\,S_-(i,\tau)-{\eta\over 2}\,S_+(i,\tau)-
\epsilon^*_+\,E(i,\tau)\ \ ,\cr
\delta S_-(i,\tau) &=+{\theta^0\over 2}
\,S_-(i,\tau) -\theta^-\,S_+(i,\tau)-
{\eta\over 2}\,S_-(i,\tau)-\epsilon^*_-\,
E(i,\tau)\ \ ,\cr
\delta S^*_+(i,\tau) &=+{\theta^0\over 2}\,
S^*_+(i,\tau) +\theta^-\,S^*_-(i,\tau)+
{\eta\over 2}\,S^*_+(i,\tau)+\epsilon_+
\,E^*(i,\tau)\ \ ,\cr
\delta S^*_-(i,\tau) &=-{\theta^0\over 2}\,
S^*_-(i,\tau) +\theta^+\,S^*_+(i,\tau)+{\eta\over
2}\,S^*_-(i,\tau)+\epsilon_-\,E^*(i,\tau)\ \ .
}\eqno(3.22)
$$
Here $\theta^0$, $\theta^\pm$ and $\eta$ are
infinitesimal real parameters associated to the bosonic
generators $J^0$, $J^\pm$ and $T$ respectively, whereas
$\epsilon_\alpha$ and $\epsilon^*_\alpha$ are infinitesimal
grassmann parameters associated to the fermionic generators
$Q_\alpha$ and $Q^\dagger_\alpha$. It is straightforward
to verify that the action (2.33) is invariant under the
transformations (3.22) if the coupling constants are
chosen as in (3.1) and the constraint $\Phi(i,\tau)=0$
is enforced.
Notice that (3.22) are valid both in the slave
fermion and in the slave boson representations,
the only difference being the statistics of the
fields. On the contrary, (3.22) are not suited for
a discussion of the supersymmetry of the $t$-$J$ model
in the slave anyon representation, since in this case
holons and spinons are anyonic {\it non-local} objects
for which a functional approach is not well-defined
(cf (2.41) and following comments). Therefore, the analysis
of the invariance of the $t$-$J$ model under the superalgebra
$SU(1|2)$ in the slave anyon representation must be
performed in the operator approach, and to this task we
turn in the following section.

\vskip 2cm
\centerline{\bf 4. The Slave Anyon Representation of
the Supersymmetric $t$-$J$ Model}
\vskip 0.7cm

In Section 2 we discussed the slave anyon representation
of the two-dimensional $t$-$J$ model, and showed that
holons and spinons become anyons of opposite statistics if
one introduces two independent Chern-Simons fields in the
bosonization procedure.
In this section we want to extend the analysis of the
supersymmetry properties presented in Section 3 to the slave
anyon representation of the  $t$-$J$ model.
To this aim, we first introduce anyonic operators by means
of a generalized Jordan-Wigner transformation [34,35].
This is given by Eq. (2.41), which, when written in terms
of oscillators (denoted by lower case letters), becomes
$$
\eqalign{
{\hat e}(i) &\equiv ~{\rm e}^{\,{\rm i}\,\nu\,
\sum\limits_{j\in \Omega}\Theta(i,j)\,\big(
e^\dagger(j)\,e(j) - 1\big)}~e(i)\ \ ,\cr
{\hat s}_\alpha(i) &\equiv ~{\rm e}^{\,-{\rm i}\,
\nu\,\sum\limits_{j\in \Omega}\Theta(i,j)\,
s_\alpha^\dagger(j)\,s_\alpha(j) }~s_\alpha(i)\ \ .
}
\eqno(4.1)
$$
Here $e(i)$ and $s_\alpha(i)$ are the holon and spinon
oscillators in the slave fermion
description, $\nu$ is the statistical parameter related to
the Chern-Simons actions as in (2.39), and $\Theta(i,j)$
is the lattice angle function. The exponential factors in
(4.1) are sometimes called disorder operators [35]. One can
prove that ${\hat e}(i)$ and ${\hat s}_\alpha (i)$ are
anyons of statistics $\nu$ and $-\nu$ respectively, and
satisfy braiding relations among themselves. To give a
rigorous discussion of these braiding relations, a detailed
analysis of the lattice angle function is necessary.
We recall that $\Theta(i,j)$ naively denotes the angle
from a base point $B$ (eventually taken to infinity) to
$i$ as seen by $j$. However, some care must be used to
define this angle on a lattice. In fact, to avoid ambiguities,
the center of the angle $\Theta(i,j)$ must be a point of the
dual lattice near $j$ and a path from $B$ to $i$ must be
chosen [26,27,28]. Once these two specifications are given,
the lattice angle function is unambiguously defined.
Different choices of the center of $\Theta(i,j)$ or
of the path along which $\Theta(i,j)$ is measured, lead
to different angle functions and hence, through (4.1),
to different anyonic oscillators characterized by
different braiding properties. This point was discussed
in great detail in [28,36] where the relation between
anyons and quantum groups was established.
Here, instead, we are interested in a different perspective,
and will choose the angle $\Theta(i,j)$ to define holon and
spinon operators in such a way that the supersymmetry
properties of the $t$-$J$ model in the slave anyon representation
become as simple as possible.

We begin by constructing the Hamiltonian of the $t$-$J$
model using the anyonic oscillators (4.1).
We just sketch its derivation without entering the details
since only elementary manipulations are involved. We start
from the action (2.37) and fix the coupling constants as
in (3.1). Then, we compute the corresponding Hamiltonian,
rewrite it in the operator formalism and modify it by
adding the constraint as we did in (3.10).  Finally,
after removing the Chern-Simons fields, we can introduce
the anyonic operators (4.1) and obtain the following
Hamiltonian
$$
H'_{tJ} = {J\over 2}\sum_{<i,j>}
\Big|{\hat s}^\dagger_\alpha(i)\,
{\hat s}_\alpha(j)-{\hat e}(j)\,
{\hat e}^\dagger(i)\Big|^2 +
\sum_{i\in\Omega}{\rm i}\,\Lambda'(i)\,\phi(i)
\eqno(4.2)
$$
where $\Lambda'(i)$ is the Lagrange
multiplier enforcing the constraint $\phi(i)=0$
to select the physical states of the $t$-$J$ model.
This constraint can be written equivalently either
with the anyonic operators (4.1) or with the
oscillators in the slave fermion representation; in fact
$$
\eqalign{
\phi(i) &= {\hat s}^\dagger_\alpha(i)\,
{\hat s}_\alpha(i)+{\hat e}^\dagger(i)\,
{\hat e}(i)-1\cr
&=s^\dagger_\alpha(i)\,s_\alpha(i)+
e^\dagger(i)\,e(i)-1\ \ .
}\eqno(4.3)
$$
This equality follows from the fact that the
disorder operators cancel exactly in the
combinations ${\hat s}_\alpha^\dagger(i)\,
{\hat s}_\alpha(i)$ and ${\hat e}^\dagger(i)
\,{\hat e}(i)$, and hence any dependence on
the statistical parameter $\nu$ and on the angle
function drops out from $\phi(i)$.

For later convenience, and in analogy with (3.14), we
introduce the operator
$$
{\hat G}(i,j) = {\hat s}^\dagger_\alpha(i)\,
{\hat s}_\alpha(j)-{\hat e}(j)\,{\hat e}^\dagger(i)
{}~~~~~~{\rm for}~~i\not=j \ \ ,
\eqno(4.4)
$$
which appears inside the absolute value of (4.2) and
is such that
$$
\Big[\phi(k)\,,\,\big|{\hat G}(i,j)\big|^2\Big]=0 \ \ .
\eqno(4.5)
$$
for all $k$, as one can easily check with some
straighforward algebra.
Notice that the holon term  of ${\hat G}(i,j)$ has
a relative minus sign with respect to the spinon term,
and that the creation operator ${\hat e}^\dagger(i)$
stands on the right of the annihilation operator
${\hat e}(j)$. Obviously the standard normal order
can be restored by exchanging ${\hat e}(j)$ and
${\hat e}^\dagger(i)$, but then, due the braiding
properties of these oscillators, a phase depending
on the relative position of $i$ and $j$ is produced
(see for example [28]).
Of course, when $\nu=0$, {\it i.e.} when the holons
are fermionic and the spinons bosonic, this phase factor
is simply a minus sign and (4.4) reduces exactly to (3.14).
Moreover, in this case the Hamiltonian (4.2) becomes equal
to that of the slave fermion representation given in (3.13),
apart from the irrelevant constant proportional to the volume
of the lattice that we have dropped.

The anyonic oscillators (4.1) and their braiding properties
crucially depend on how one chooses the angle function in the
disorder operators [28]; on the contrary the Hamiltonian (4.2)
is independent of such a choice. To prove this, let us consider
the operator
$$
{\hat G}'(i,j) = {\hat s}'^\dagger_\alpha(i)\,
{\hat s}'_\alpha(j)-{\hat e}'(j)\,{\hat e}'^\dagger(i)  \ \,
\eqno(4.6)
$$
where the primed anyonic oscillators are defined as in
(4.1) but with a different angle function denoted by
$\Theta'(i,j)$. Expressing these primed oscillators in
terms of the unprimed ones, (4.6) becomes
$$
\eqalign{
{\hat G}'(i,j) =&~ {\hat s}^\dagger_\alpha(i)~
q^{~\sum\limits_{k\in\Omega}\big[
\chi(i,k)-\chi(j,k)\big]\,s^\dagger_\beta(k)\,s_\beta(k)}~
{\hat s}_\alpha(j) \cr
&-
q^{~\sum\limits_{k\in\Omega}
\chi(j,k)\,\big(e^\dagger(k)\,e(k)-1\big)}~{\hat e}(j)\,
{\hat e}^\dagger(i)
{}~q^{\,-\sum\limits_{k\in\Omega}
\chi(i,k)\,\big(e^\dagger(k)\,e(k)-1\big)}}
\eqno(4.7)
$$
where
$$
q={\rm e}^{\,{\rm i}\,\pi\,\nu}\ \ ,
\eqno(4.8)
$$
and
$$
\chi(i,j) = {1\over \pi}~\Big[\Theta'(i,j)-{\Theta}(i,j)\Big]\ \ .
\eqno(4.9)
$$
If we rewrite the exponents in the holon term using (4.3),
and then rearrange the factors, we get
$$
\eqalign{
{\hat G}'(i,j) = &q^{\Big\{\chi(j,i)+
\sum\limits_{k\in\Omega}\big[
\chi(i,k)-\chi(j,k)\big]\,s^\dagger_\beta(k)\,
s_\beta(k)\Big\}}~\Bigg\{
q^{-\chi(i,i)}~{\hat s}^\dagger_\alpha(i)\,
{\hat s}_\alpha(j) \cr
&~~~~~~~~~~~~~~~~~-q^{-\chi(j,j)}~{\hat e}(j)\,
{\hat e}^\dagger(i)~
q^{\,-\sum\limits_{k\in\Omega}\big[
\chi(i,k)-\chi(j,k)\big]\,\phi(k)}\Bigg\} \ \ .
}\eqno(4.10)
$$
{}From the definition (4.9), it is clear that
$$
\chi(i,i)=\chi(j,j)=\chi \ \ .
\eqno(4.11)
$$
If we take into account the constraint $\phi(i)=0$
for all $i\in\Omega$, the right hand side of (4.10)
simplifies considerably and one gets
$$
{\hat G}'(i,j) = q^{\Big\{-\chi+\chi(j,i)+
\sum\limits_{k\in\Omega}\big[
\chi(i,k)-\chi(j,k)\big]\,
s^\dagger_\beta(k)\,s_\beta(k)\Big\}}~~
{\hat G}(i,j)\ \ .
\eqno(4.12)
$$
Therefore, on the physical states of the $t$-$J$ model,
${\hat G}'(i,j)$ and ${\hat G}(i,j)$ simply differ by a
unitary operator, and thus
$$
\Big|{\hat G}'(i,j)\Big|^2={\hat G}'^\dagger(i,j)\,
{\hat G}'(i,j) = {\hat G}^\dagger(i,j)\,{\hat G}(i,j)=
\Big|{\hat G}(i,j)\Big|^2
\eqno(4.13)
$$
for all $i$ and $j$. This equality implies that the
Hamiltonian (4.2) does not depend on the details of the
angle function. This property should not be totally
unexpected, since different choices of the angle function
simply correspond to different gauges for the Chern-Simons
fields used in the bosonization procedure.

We now come to the slave anyon representation of the
superalgebra $SU(1|2)$.
As far as the generators $J^\pm$, $J^0$ and $T$ are
concerned, there is no difference with respect to the
case discussed in Section 3, and Eqs. (3.6a-b) and
(3.7a-b) hold also in the anyonic representation,
apart from some obvious notational changes. Things
are different instead for the supersymmetry generators
 $Q_\alpha$ and $Q_\alpha^\dagger$. In fact, since these
are fermionic, they are ``sensitive'' to the
statistics of the operators on which they act.
For example, in the slave fermion representation
$Q_\alpha$ and $Q_\alpha^\dagger$ close
anticommutators with $e$ and $e^\dagger$, but in the
slave boson representation they close commutators, as
one can see from Eqs. (3.7c) and (3.21).
Therefore, for intermediate statistics it is natural to
expect a generalization of these equations.

The slave anyon representation of the supercharges can be
written in analogy with (3.6c);
in fact one has
$$
\eqalign{
Q_\alpha &= \sum_{i\in\Omega}{\hat Q}_\alpha(i) =
\sum_{i\in\Omega}{\hat e}^\dagger(i)
\,{\hat s}_\alpha(i) \ \ ,\cr
Q^\dagger_\alpha &= \sum_{i\in\Omega}
{\hat Q}^\dagger_\alpha(i) = \sum_{i\in\Omega}
{\hat e}(i) \,{\hat s}_\alpha^\dagger(i)\ \ .}
\eqno(4.14)
$$
With simple algebra using (4.1) and (2.12), one can check
that $Q_\alpha$ and $Q_\alpha^\dagger$ in (4.14) close the
proper anticommutators of $SU(1|2)$, and also that the
densities ${\hat Q}_\alpha(i)$ and
 ${\hat Q}^\dagger_\alpha(i)$, while being fermions with
respect to each other, are anyons with respect to the
constituent holons and spinons. Indeed, ${\hat Q}_\alpha(i)$
and ${\hat Q}^\dagger_\alpha(i)$ close anticommutators
among themselves, but obey braiding relations with
${\hat e}(i)$ and ${\hat s}_\alpha(i)$.
Let us now analyze if the supercharges (4.14)
depend on the choice of the angle function in the
anyonic oscillators. To this purpose, we consider
the density
$$
{\hat Q}'_\alpha(i) = {\hat e}'^\dagger(i) \,
{\hat s}'_\alpha(i)
\eqno(4.15)
$$
where the primed oscillators are those
introduced earlier. Expressing the latter ones in
terms of the unprimed oscillators (4.1), we get
$$
{\hat Q}'_\alpha(i) = {\hat e}^\dagger(i)~
q^{\,-\sum\limits_{j\in \Omega}\chi(i,j)\,
\phi(j) }~{\hat s}_\alpha(i) \ \ .
\eqno(4.16)
$$
Bringing the $q$-factor to the far right and
taking into account (4.11), we finally obtain
$$
{\hat Q}'_\alpha(i)  = q^\chi~{\hat Q}_\alpha(i)~
q^{\,-\sum\limits_{j\in\Omega}
\chi(i,j)\,\phi(j)} \ \ .
\eqno(4.17)
$$
This equation implies that on the physical states of
the $t$-$J$ model, which satisfy $\phi(i)=0$ for all
$i\in\Omega$, the two supersymmetry densities
${\hat Q}'_\alpha(i)$ and
${\hat Q}_\alpha(i)$ differ only by a phase factor
independent of $i$. Thus, on the physical states also
the corresponding supersymmetry generators simply
differ by a constant phase:
$$
Q'_\alpha = \sum_{i\in\Omega}{\hat Q}'_\alpha(i) =
q^\chi~\sum_{i\in\Omega}{\hat Q}_\alpha(i) =
q^\chi~Q_\alpha \ \ ,
\eqno(4.18)
$$
and hence they close the same superalgebra (3.3).
In fact, the (anti)commutation relations of $SU(1|2)$
remain unchanged if the supersymmetry charges are
rescaled by a phase factor. Moreover, it is clear
that if $Q_\alpha$ commutes with the Hamiltonian,
so does $Q'_\alpha$.

The result contained in Eqs. (4.13) and (4.18) is
quite important: All possible choices of angle function
for the anyonic oscillators are equivalent both in the
Hamiltonian and in the supersymmetry charges. Therefore,
we are free to choose those angle functions that are
most convenient for our purposes and that simplify most
our calculations. Actually, in discussing the
supersymmetry properties of the slave anyon
representation of the $t$-$J$ model we find very
convenient to introduce two angle functions that we
call horizontal and vertical, and correspondingly
define two kinds of anyonic oscillators that we also
call horizontal (or of type ${\cal H}$) and vertical
(or of type ${\cal V}$). The reason for this
terminology will be evident in a moment.

For the type ${\cal H}$ anyons, the angle function
$\Theta_{\cal H}(i,j)$ is defined as follows: First,
for any point $i\in\Omega$ we consider a straight lattice
path ${\cal C}_{\cal H}(i)$ parallel to the $x$-axis
from the base point $B$ (taken to infinity in the
direction of the positive $x$-axis) to the point $i$;
then we take the dual point
$$
j^*\equiv\Big(j_x+{\varepsilon\over 2}~,~j_y+
{\varepsilon\over 2}\Big)
\eqno(4.19)
$$
where $j_x$ and $j_y$ are the coordinates of
$j\in\Omega$, and $\varepsilon\le 1$ is the spacing
of a new lattice $\Omega'$ in which $\Omega$ is embedded
(eventually $\varepsilon$ can be taken to zero, see [28]).
The point $j^*$ belongs to the lattice dual to
$\Omega'$, and can also be viewed as the center of
the elementary plaquette of $\Omega'$ whose lower
left corner is $j$.
Finally, we define $\Theta_{\cal H}(i,j)$ as the angle
between $B$ and $i$ measured from $j^*$ along the path
${\cal C}_{\cal H}(i)$. This is represented in Fig. 1.

Using this horizontal angle in the Jordan-Wigner
transformation (4.1) we get the anyons of type
${\cal H}$, which from now on will be simply denoted
by the symbols ${\hat e}(i)$, ${\hat e}^\dagger(i)$,
${\hat s}_\alpha(i)$ and ${\hat s}^\dagger(i)$ with
no further specifications.
These operators satisfy braiding relations among themselves
that can be easily obtained using the (anti)commutators
(2.12) and the properties of the angle function
$\Theta_{\cal H}(i,j)$. For example, if $i$ and $j$
are two horizontal nearest neighbor points, {\it viz.}
$j=i+{\hat 1}$, one can show that in the limit
$\varepsilon\to 0$
$$
\eqalignno{
{\hat e}(i)~{\hat e}(j)&+q~ {\hat e}(j)~
{\hat e}(i) =0 \ \ , &(4.20{\rm a})\cr
{\hat e}(j)~{\hat e}^\dagger(i)&+q~
{\hat e}^\dagger(i)~ {\hat e}(j) =0 \ \ ,
&(4.20{\rm b})\cr
{\hat s}_\alpha(i)~{\hat s}_\beta(j) &-q^{-1}~
{\hat s}_\beta(j) ~ {\hat s}_\alpha(i) =0 \ \ ,
 &(4.20{\rm c})\cr
{\hat s}_\alpha(j)~{\hat s}^\dagger_\beta(i) &-q^{-1}~
{\hat s}^\dagger_\beta(i) ~ {\hat s}_\alpha(j) =0
&(4.20{\rm d})
}
$$
where $q$ is given in (4.8).
We refer the reader to [28] where these braiding
relations are discussed and derived in full generality.
Here we simply point out that there are no phases in
the relations between operators in the same point
which indeed are
$$
\eqalignno{
{\hat e}(i)~
{\hat e}^\dagger(i)&+
{\hat e}^\dagger(i)~
{\hat e}(i)=1 \ \ , &(4.21{\rm a})\cr
{\hat s}_\alpha(i)~
{\hat s}^\dagger_\beta(i)&-
{\hat s}^\dagger_\beta(i)~
{\hat s}_\alpha(i)=\delta_{\alpha\beta}\ \ . &(4.21{\rm b})
}
$$
It is also important to realize that ${\hat e}$ and
${\hat s}_\alpha$ commute among themselves; for
example
$$
{\hat e}(i)~{\hat s}_\alpha(j)-{\hat s}_\alpha(j)~
{\hat e}(i) = 0
\eqno(4.22)
$$
for all $i$ and $j$. From Eqs. (4.20), (4.21) and (4.22),
it is straightforward to obtain the braiding relations between
the supercharge densities ${\hat Q}_\alpha(i)$ and
${\hat Q}^\dagger_\alpha(i)$ and the anyonic holon and
spinon operators. These braiding relations generalize
Eqs. (3.7c) and (3.21) to the case of arbitrary statistics,
and provide an anyonic linear realization of the $N=2$
supersymmetry.

Let us now come to the anyons of type ${\cal V}$, which
are characterized by the vertical angle function
$\Theta_{\cal V}(i,j)$. The latter is defined as
$\Theta_{\cal H}(i,j)$ but with a rotation of $\pi/2$
in a counterclockwise way. More precisely, for any
point $i\in \Omega$ we consider a straight lattice
path ${\cal C}_{\cal V}(i)$ parallel to the $y$-axis
from the base point $B'$ (taken to infinity in the
direction of the positive $y$-axis) to $i$; then given
$j\in \Omega$, we take the dual point
$$
{}^*\!j\equiv\Big(j_x-{\varepsilon\over 2}~,~j_y+
{\varepsilon\over 2}\Big)
\eqno(4.23)
$$
which is the center of the elementary plaquette of
$\Omega'$ whose lower right corner is $j$, and define
$\Theta_{\cal V}(i,j)$ as the angle between $B$ and $i$
measured from  ${}^*\!j$ along the path
${\cal C}_{\cal V}(i)$. This is represented
in Fig. 2.

The anyonic oscillators of type ${\cal V}$ are obtained
by using the vertical angle function $\Theta_{\cal V}(i,j)$
in (4.1); they satisfy braiding relations among themselves
that can be easily determined from the oscillator algebra
(2.12), and are formally the same as those satisfied by the
horizontal anyons (indeed, Eqs. (4.20) still hold with
$j=i+{\hat 2}$).

These two types of anyonic oscillators are very helpful in
simplifying our calculations. In fact, if we distinguish
between horizontal and vertical lattice bonds and
correspondingly introduce horizontal and vertical
anyons,
the Hamiltonian (4.2) can be conveniently rewritten
as follows
$$
\eqalign{
H'_{tJ}&= H_{\cal H} + H_{\cal V} +
\sum_{i\in\Omega}{\rm i}\,\Lambda'(i)\,\phi(i) \cr
&={J\over 2}\!\sum_{<i,j>_{\cal H}}
\!\Big|G^{\cal H}(i,j)\Big|^2 +{J\over 2}\!
\sum_{<i,j>_{\cal V}}
\!\Big|G^{\cal V}(i,j)\Big|^2 + \sum_{i\in\Omega}
{\rm i}\,\Lambda'(i)\,\phi(i)
}
\eqno(4.24)
$$
where
$$
G^{\cal H}(i,j)={\hat s}^\dagger_\alpha(i)\,
{\hat s}_\alpha(j)-{\hat e}(j)\,{\hat e}^\dagger(i)
\eqno(4.25)
$$
and $G^{\cal V}(i,j)$ is the same but with vertical
anyons.
The horizontal Hamiltonian $H_{\cal H}$ contains only
the interactions between horizontal nearest-neighbor
 points denoted by $<i,j>_{\cal H}$, whereas the vertical
Hamiltonian $H_{\cal V}$ contains only the contributions
from vertical nearest-neighbor points
denoted by $<i,j>_{\cal V}$.
In some extended sense, $H_{\cal H}$ and
 $H_{\cal V}$ can be considered as one-dimensional
Hamiltonians corresponding respectively to the rows
and the columns of the original two-dimensional square lattice.

Actually, in writing the Hamiltonian (4.24), we
could have used only one kind of anyonic
oscillators, for instance only horizontal ones.
But in such a case, the term corresponding to
vertical nearest neighbor interactions, $G^{\cal V}(i,j)$,
would have been more complicated and would have contained
extra non-local operators, as shown in [27] for a
different model. However, as we discussed before, these
extra non-local operators can be written simply as
$q$ raised to suitable combinations of the holon and
spinon densities, and, after using the constraint
$\phi(i)=0$, they can be reabsorbed into an overall
unitary operator which drops out in taking the
absolute value (see (4.12) and (4.13)).
Finally, we remark that when $\nu=0$ holons and spinons
become fermions and bosons respectively, and the
distinction between horizontal and vertical operators
ceases to exist. In this case the Hamiltonians $H_{\cal H}$
and $H_{\cal V}$ in (4.24) can be combined, and together
they reproduce exactly the Hamiltonian (3.13) of the
supersymmetric $t$-$J$ model in the slave fermion
representation, up to an irrelevant additive
constant that we have dropped.

Let us now discuss the supersymmetry properties the
$t$-$J$ model in the slave anyon representation. To
this purpose, it is convenient to start from the
form (4.24) of the Hamiltonian and assume that the
supersymmetry generators $Q_\alpha$ and $Q_\alpha^\dagger$
are made of anyonic oscillators of type ${\cal H}$.
Since we are free to choose any angle function due to
the relation (4.18), there is no loss of generality in
making this assumption. Moreover, to avoid ambiguities
we denote the supercharge densities by the symbols
${\hat Q}_\alpha^{\cal H}(i)$ and
${\hat Q}_\alpha^{\dagger\,{\cal H}}(i)$.  From
the braiding relations (4.20) and the (anti)commutators
(4.21) and (4.22), it follows that
$$
\Big[{\hat Q}_\alpha^{\cal H}(k)\,,
\,{\hat G}^{\cal H}(i,j)\Big] =
\Big[{\hat Q}_\alpha^{\cal H}(k)\,,
\,{\hat G}^{\dagger\,{\cal H}}(i,j)\Big]= 0
{}~~~~{\rm for}~~ k\not=i,j\ \ ,
\eqno(4.26)
$$
and
$$
\eqalign{
\Big({\hat Q}_\alpha^{\cal H}(i)+
{\hat Q}_\alpha^{\cal H}(j)\Big)\,
{\hat G}^{\cal H}(i,j)&= q^{-1}~
{\hat G}^{\cal H}(i,j)\,
\Big({\hat Q}_\alpha^{\cal H}(i)+
{\hat Q}_\alpha^{\cal H}(j)\Big)\ \ ,\cr
\Big({\hat Q}_\alpha^{\cal H}(i)+
{\hat Q}_\alpha^{\cal H}(j)\Big)\,
{\hat G}^{\dagger\,{\cal H}}(i,j)&=
q~{\hat G}^{\dagger\,{\cal H}}(i,j)\,
\Big({\hat Q}_\alpha^{\cal H}(i)+
{\hat Q}_\alpha^{\cal H}(j)\Big) \ \ .}
\eqno(4.27)
$$
Of course the corresponding equations involving
the density ${\hat Q}_\alpha^{\dagger\,{\cal H}}(i)$
can be simply obtained from (4.26) and (4.27) by
hermitean conjugation. These equations imply that
$$
\Big[Q_\alpha\,,\, \big|{\hat G}^{\cal H}(i,j)
\big|^2\Big] =
\Big[Q^\dagger_\alpha\,,\, \big|{\hat G}^{\cal H}(i,j)
\big|^2\Big] = 0 \ \ ,
\eqno(4.28)
$$
and hence the horizontal Hamiltonian $H_{\cal H}$ is
invariant under supersymmetry.

To prove that also the vertical Hamiltonian
commutes with $Q_\alpha$ and $Q_\alpha^\dagger$,
we first observe that Eqs. (4.26) and (4.27) can be
translated into the corresponding ones for vertical
anyons by simply replacing everywhere the superscript
${\cal H}$ with ${\cal V}$. Then we make use of (4.18)
and its hermitean conjugate to rewrite ${\hat Q}_\alpha$
and
${\hat Q}_\alpha^\dagger$ in terms of vertical
anyons, and conclude that
$$
\Big[Q_\alpha\,,\, \big|{\hat G}^{\cal V}(i,j)
\big|^2\Big] =
\Big[Q^\dagger_\alpha\,,\,
\big|{\hat G}^{\cal V}(i,j)\big|^2\Big] = 0 \ \ .
\eqno(4.29)
$$
Finally, it is straightforward to check that
the supercharges commute with the constraint
operator $\phi(i)$. This completes the proof
of the invariance of the whole Hamiltonian
(4.24) under the $N=2$ supersymmetry.

In Appendix A we will show that this same result
can be obtained by a different definition of the angle
function in the anyonic oscillators forming the
supersymmetry generators
${\hat Q}_\alpha$ and
${\hat Q}_\alpha^\dagger$. This alternative method is
 based on a more intensive use of the constraint
$\phi(i)=0$, but has the advantage of making ${\hat G}(i,j)$,
and not only its modulus square, invariant under $SU(1|2)$.

\vskip 2cm
\centerline{\bf 5. Conclusions}
\vskip 0.7cm

In this paper we have studied the formal properties of the
$t$-$J$ model in two dimensions using the formalism of the
slave operators, and showed that, by means of a generalized abelian
bosonization with two independent Chern-Simons fields,
holons and spinons may acquire arbitrary complementary
statistics in such a way that the electron remains
a fermion. We would like to stress that
the factorization of the electron in an antiholon and
a spinon is a formal {\it Ansatz},
and only a detailed analysis at the dynamical level could
shed some light on its correctness.

The $t$-$J$ model at $J=2\,t$ is characterized by the invariance
of its Hamiltonian under the superalgebra $SU(1|2)$, which is linearly
realized on holons and spinons. In the case of anyonic statistics, the
supercharge densities of $SU(1|2)$ have braiding properties
with the holon and
spinon operators, which generalize the standard (anti)commutation
relations with fermions and bosons. These supersymmetry
properties of
the $t$-$J$ model at $J=2\,t$ could play an important role in the
analysis of the dynamical aspects of the spin-charge
separation, which
certainly deserves to be investigated further.

\vskip 3cm
\centerline{\bf Acknowledgements}
\vskip 0.7cm
\noindent
We would like to thank Fabian Essler and Vladimir
Korepin for very inspiring discussions
during the early stages of this work.
\vfill
\eject
\centerline{\bf Appendix A}
\vskip 0.7cm
In this Appendix we propose two new definitions for
the angle function in the anyonic oscillators forming
the supercharges ${\hat Q}_\alpha$ and
${\hat Q}_\alpha^\dagger$ that simplify considerably
the discussion of the supersymmetry properties of the
$t$-$J$ model in the slave anyon representation. In fact
with this choice the operator ${\hat G}(i,j)$, and not
only $\big|{\hat G}(i,j)\big|^2$, will be supersymmetric.
We denote these new angle functions by
${\tilde \Theta}_{\cal H}(i,j)$ and
${\tilde \Theta}_{\cal V}(i,j)$ corresponding
respectively to the horizontal and vertical anyons.
The function ${\tilde \Theta}_{\cal H}(i,j)$ is
defined as follows: For any point $i\in\Omega$,
we choose the horizontal path ${\cal C}_{\cal H}(i)$
from the base point $B$ to $i$, and consider the dual point
$$
{}_*j\equiv\Big(j_x-{\varepsilon\over 2}~,~j_y-
{\varepsilon\over 2}\Big)
\eqno({\rm A}.1)
$$
which is the center of the elementary plaquette of
$\Omega'$ whose upper right corner is $j$. Then we
define ${\tilde \Theta}_{\cal H}(i,j)$ as the angle
between $B$ and $i$ measured from ${}_*j$ along the
path ${\cal C}_{\cal H}(i)$. This is represented in Fig. 3.

Using this twiddled horizontal angle in the Jordan-Wigner
transformation (4.1) we get twiddled anyons of type
${\cal H}$, which, from now on, we simply denote by
the symbols ${\tilde {\hat e}}(i)$,
${\tilde {\hat e}}^\dagger(i)$, ${\tilde {\hat s}}_\alpha(i)$
and ${\tilde {\hat s}}^\dagger(i)$ with no further
specifications.
These twiddled horizontal oscillators are used to
define a new anyonic representation of the supersymmetry
generators. In fact, we can posit
$$
\eqalign{
Q_\alpha &= \sum_{i\in\Omega}{\tilde
{\hat Q}}_\alpha(i) = \sum_{i\in\Omega}
{\tilde{\hat e}}^\dagger(i)
\,{\tilde {\hat s}}_\alpha(i) \ \ ,\cr
Q^\dagger_\alpha &= \sum_{i\in\Omega}{\tilde
{\hat Q}}^\dagger_\alpha(i) = \sum_{i\in\Omega}
{\tilde {\hat e}}(i) \,
{\tilde {\hat s}}_\alpha^\dagger(i)\ \ .}
\eqno({\rm A}.2)
$$
As discussed in Section 4, these
supercharges are equivalent to those
defined in (4.14), since they simply differ in the
definition of the angle function in the anyonic
oscillators (cf (4.18)).

We now study the braiding relations between the
supersymmetry generators (A.2) and the Hamiltonian
(4.24) which is made out of anyons
of type ${\cal H}$ and type ${\cal V}$. To this aim it
is therefore necessary to establish the precise
connection between ${\tilde \Theta}_{\cal H}(i,j)$
and the horizontal and vertical angle functions defined
in Section 4. To avoid repetitions, we limit our
discussion to the horizontal case, but of course
similar considerations can be made also in the vertical case.

{}From the definitions of $\Theta_{\cal H}(i,j)$ and
${\tilde \Theta}_{\cal H}(i,j)$, one can derive the
following identity
$$
\Theta_{\cal H}(i,j) - {\tilde \Theta}_{\cal H}(j,i)=
\left\{\matrix{
 \pi ~~~~~~~~~~{\rm for}~~~i_y>j_y \ \ , \cr
 -\pi ~~~~~~~~{\rm for}~~~i_y\leq j_y \ \ ,
\cr}\right.
\eqno({\rm A}.3)
$$
which holds for any value of $\varepsilon$ and not only
for $\varepsilon\to 0$. The derivation of this identity is
completely similar to that discussed in detail in [27,28],
and so we do not reproduce it here. Notice that the case $i=j$
is included in this equation since
$$
\Theta_{\cal H}(i,i)=-{3\pi\over 4}~~~~~~,~~~~~~
{\tilde \Theta}_{\cal H}(i,i)={\pi\over 4}
\eqno({\rm A}.4)
$$
for any $i\in\Omega$.

Let us now consider two anyonic holons, one in the
horizontal representation
and one in the twiddled representation. With simple
algebra one can prove that
$$
{\hat e}(i)~ {\tilde {\hat e}}(j) = -
{\rm e}^{-{\rm i}\,\nu
\big[\Theta_{\cal H}(i,j)-{\tilde \Theta}_{\cal H}(j,i)\big]}~
{\tilde {\hat e}}(j)~{\hat e}(i)\ \ .
\eqno({\rm A}.5{\rm a})
$$
Using the identity (A.3), this equation becomes
$$
{\hat e}(i)~ {\tilde {\hat e}}(j)=
\left\{\matrix{
 -q^{-1}~{\tilde {\hat e}}(j)~{\hat e}(i) ~~~~~~
{\rm for}~~~i_y>j_y \ \ , \cr
 {}\cr
 -q~{\tilde {\hat e}}(j)~{\hat e}(i)~~~~~~~~~
{\rm for}~~~i_y\leq j_y \ \ .
\cr}\right.
\eqno({\rm A}.5{\rm b})
$$
It is convenient to introduce the function
$\sigma$, defined on the integers as
$$
\sigma(m) = \theta(m) - \theta(-m) +\delta_{m,0} =
\left\{\matrix{
1~~~~~~~~{\rm for}~~~m\geq 0\ \ , \cr
-1 ~~~~~~{\rm for}~~~m<0 \ \ ,\cr}\right.
\eqno({\rm A}.6{\rm b})
$$
where $\theta(m)$ is the standard Heaviside step function
$$
\theta(m) = \left\{\matrix{
1 ~~~~~~~~~~{\rm for}~~~m> 0 \ \ , \cr
{1\over 2}  ~~~~~~~~~~{\rm for}~~~m= 0 \ \ , \cr
0 ~~~~~~~~~~{\rm for}~~~m<0 \ \ .
\cr}\right.
\eqno({\rm A}.6{\rm b})
$$
Then, we can rewrite (A.5b) as
$$
{\hat e}(i)~ {\tilde {\hat e}}(j) + q^{\sigma(j_y-i_y)}~
{\tilde {\hat e}}(j)~{\hat e}(i) = 0\ \ ,
\eqno({\rm A}.7)
$$
which is an example of the braiding relations we are
looking for.
We now consider how creation and annihilation operators
braid among themselves. For $i\not=j$ we get
$$
{\hat e}(i)~{\tilde {\hat e}}^\dagger(j)=\left\{\matrix{
 -q~{\tilde {\hat e}}^\dagger(j)~{\hat e}(i) ~~~~~~~~
{\rm for}~~i_y>j_y  \cr
 {}\cr
  -q^{-1}~{\tilde {\hat e}}^\dagger(j)~{\hat e}(i)~~~~~~
{\rm for}~~i_y\leq j_y \cr}\right.
\eqno({\rm A}.8)
$$
where again we have exploited the identity (A.3). If
$i=j$ instead, using
(A.4) and taking into account the inhomogeneous term in the
anticommutation relations  (2.12a) at the same point,
we obtain
$$
{\hat e}(i)~{\tilde {\hat e}}^\dagger(i) +
 q^{-1}~{\tilde {\hat e}}^\dagger(i)~{\hat e}(i)
= q^{~\sum\limits_{j\in\Omega}
\chi(i,j)\,\big(e^\dagger(j)\,e(j)-1\big)}
\eqno({\rm A}.9)
$$
where
$$
\chi(i,j) = {1\over \pi}~\Big[\Theta_{\cal H}(i,j)-
{\tilde \Theta}_{\cal H}(i,j)\Big]\ \ .
\eqno({\rm A}.10)
$$
Notice that there is a braiding phase even in the
relation at the same point, in contrast to (4.21a)
which is a standard anticommutator.
We can easily combine Eqs. (A.8) and (A.9) into a
single formula as follows
$$
{\hat e}(i)~{\tilde {\hat e}}^\dagger(j) +
q^{-\sigma(j_y-i_y)}~
{\tilde {\hat e}}^\dagger(j)~{\hat e}(i) =
\delta(i,j)~q^{~\sum\limits_{k\in\Omega}
\chi(i,k)\,\big(e^\dagger(k)\,e(k)-1\big)}\ \ .
\eqno({\rm A}.11)
$$
Eqs. (A.7), (A.11) and their hermitean conjugates
form the complete set of braiding relations between
two anyonic holon operators in the horizontal
and twiddled representations.

It is interesting to consider these relations in
the limit $\varepsilon\to 0$ which is part of the
continuum limit (we recall that $\varepsilon$ is the
spacing of the lattice $\Omega'$ where $\Omega$ is
embedded, and is used to determine the centers from
which the angles are measured, see for instance (4.2) and (A.1)).
The braiding relation (A.7) remains unchanged when
$\varepsilon \to 0$, whereas (A.11) simplifies a little,
since the function $\chi(i,k)$ appearing in the right hand
side can be computed explicitly. In fact,
from the definitions of $\Theta_{\cal H}(i,j)$ and
${\tilde \Theta}_{\cal H}(i,j)$, it is not difficult
to show that, when $\varepsilon\to 0$,
$$
\eqalign{
\Theta_{\cal H}(i,j) &= {\tilde \Theta}_{\cal H}(i,j)
{}~~~~~~~~~~~~{\rm for}~~~i_y\not=j_y \ \ ,\cr
\Theta_{\cal H}(i,j) &= -\pi~~,~~{\tilde \Theta}_{\cal H}(i,j)=
\pi ~~~~~~~~{\rm for}~~~i_y=j_y~,~ i_x<j_y \ \ ,\cr
\Theta_{\cal H}(i,j) &= {\tilde \Theta}_{\cal H}(i,j)= 0
{}~~~~~~~~~~~~~~~~{\rm for}~~~i_y=j_y~,~i_x>j_x  \ \ .\cr
}
\eqno({\rm A}.12)
$$
Using these values and those in (A.4), which hold for
any $\varepsilon$, we get
$$
\chi(i,j) = -2\,\delta_{i_y,j_y}\,\theta(j_x-i_x)
{}~~~~~~{\rm for}~~\varepsilon\to 0
\eqno({\rm A}.13)
$$
where $\theta$ is the step function (A.6b). Then,
when $\varepsilon \to 0$ we can rewrite the right hand
side of (A.12) and get
$$
{\hat e}(i)~{\tilde {\hat e}}^\dagger(j) +
q^{-\sigma(j_y-i_y)}~{\tilde {\hat e}}^\dagger(j)~
{\hat e}(i) = \delta(i,j)~q^{-2\sum\limits_{k_x}
\theta(k_x-i_x)\,\big(e^\dagger(k)\,e(k)-1\big)} \ \ .
\eqno({\rm A}.14)
$$
We remark that it is much easier to guess a continuum
limit of Eqs. (A.7) and (A.14) than of the analogous
braiding relations involving only anyons of one kind [28].

What we have done so far with the holons, can be
repeated with the spinons. Of course, since in the
slave anyon representation the spinons are boson
based , while the holons are fermion based, there are
some sign changes; moreover, since the statistics of the
spinons is opposite to that of the holons, $q$ has to be
changed into $q^{-1}$.
Keeping in mind these simple modifications, we can
immediately translate all braiding relations of the
holons into those of the spinons. For example, the
spinon version of (A.7) is
$$
{\hat s}_\alpha(i)~ {\tilde {\hat s}}_\beta(j) -
q^{-\sigma(j_y-i_y)}~
{\tilde {\hat s}}_\beta(j)~{\hat s}_\alpha(i) =
0\ \ .
\eqno({\rm A}.15)
$$
The other braiding relations of the spinon
operators can be obtained in a similar way from (A.11).
Moreover it is clear that holons and spinons commute with
each other, being independent operators (cf (4.22)).

After this detailed analisys of the braiding relations,
we discuss the supersymmetry properties of the Hamiltonian
(4.24) using the representation (A.2) of the supercharges.
Let us first consider ${\tilde {\hat Q}}_\alpha(i)$ and
the horizontal nearest neighbor interaction $G^{\cal H}(i,j)$
defined
in (4.25) for $j=i+{\hat 1}$ (which implies in particular
that $i_y=j_y$). Then, it is easy to prove that
$$
\Big[{\tilde {\hat Q}}_\alpha(k)\,,
\,G^{\cal H}(i,j)\Big] = 0
\eqno({\rm A}.16)
$$
for any $k\not=i,j$. The commutation relations of
$G^{\cal H}(i,j)$ with the supercharge densities in
$i$ and $j$ are more subtle. Due to the hard-core
condition obeyed by the holon and spinon operators
in the slave anyon representation, one immediately
 proves that
$$
\Big[{\tilde {\hat Q}}_\alpha(i)\,,\,
{\hat e}(j)\,{\hat e}^\dagger(i)\Big] =
\Big[{\tilde {\hat Q}}_\alpha(j)\,,\,
{\hat s}^\dagger_\alpha(i)\,
{\hat s}_\alpha(j)\Big] = 0
\eqno({\rm A}.17)
$$
On the contrary, after some algebra
involving the use of (A.7) and (A.11), we get
$$
\eqalign{
\Big[{\tilde {\hat Q}}_\alpha(j)\,,\,
{\hat e}(j)\,{\hat e}^\dagger(i)\Big] &=
q^{\,\chi(j,i)+1}~{\hat e}^\dagger(i)~
q^{\,\sum\limits_{k\in\Omega}
\chi(j,k)\,\big(e^\dagger(k)\,e(k)-1\big)}~
{\tilde {\hat s}}_\alpha(j)\cr
&=q^{\,\chi(j,i)+1}~{\hat e}^\dagger(i)~
q^{\,\sum\limits_{k\in\Omega}
\chi(j,k)\,\phi(k)}~{\hat s}_\alpha(j)
}\eqno({\rm A}.18)
$$
where in the last line we have transformed the
spinon oscillator from the twiddled representation
to the horizontal one, and have combined all exponents
to reproduce the constraint operator $\phi(k)$.
Similarly, one can prove that
$$
\eqalign{
\Big[{\tilde {\hat Q}}_\alpha(i)\,,\,
{\hat s}^\dagger_\alpha(i)\,{\hat s}_\alpha(j)\Big] &= {\tilde {\hat
e}}^\dagger(i)~
q^{\,\sum\limits_{k\in\Omega}\sum\limits_{\beta=\pm1}
\chi(i,k)\,
s^\dagger_\beta(k)\,s_\beta(k)}~{\hat s}_\alpha(j)\cr
&={\hat e}^\dagger(i)~q^{\,\sum\limits_{k\in\Omega}
\chi(i,k)\,\phi(k)}~{\hat s}_\alpha(j)
}\eqno({\rm A}.19)
$$
where in the last line the holon oscillator has been
transfomed from the twiddled to the horizontal
representation, and again all exponents have been
combined to yield the constraint operator. Putting together
the last three equations and rearranging the right hand
sides of (A.18) and (A.19) to bring the factors with the
constraint operators to the far right, one gets
$$
\eqalign{
\Big[Q_\alpha\,,\,G^{\cal H}(i,j)\Big] &=
\sum_{k\in \Omega}
\Big[{\tilde {\hat Q}}_\alpha(k)\,,
\,G^{\cal H}(i,j)\Big]\cr
&={\hat e}^\dagger(i)\,{\hat s}_\alpha(j)~
\Bigg\{ q^{\,-\chi(i,j)+
\sum\limits_{k\in\Omega}
\chi(i,k)\,\phi(k) }\cr
&~~~~~~~~~~~~~~~~~~~~~~~~-
q^{\,\chi(j,i)+1-\chi(j,j)+
\sum\limits_{k\in\Omega}
\chi(j,k)\,\phi(k)}\Bigg\} \ \ .
}\eqno({\rm A}.20)
$$
{}From the definition of the function
$\chi$ in (A.10), it follows that for $j=i+{\hat 1}$
$$
\chi(j,i) + 1 - \chi(j,j) = -\chi(i,j) \ \ .
$$
Therefore, the numerical factors in the two terms
inside the curly brackets of (A.20) are the same and
can be factored out, leaving only exponentials of the
constraint operator.
On the physical states of the $t$-$J$ model where the
constraint $\phi(i)=0$ is identically satisfied, each
exponential becomes 1, and thus we can conclude that
$$
\Big[Q_\alpha\,,\,G^{\cal H}(i,j)\Big] = 0\ \ .
\eqno({\rm A}.21{\rm a})
$$
Similarly, one can show that
$$
\Big[Q^\dagger_\alpha\,,\,G^{\cal H}(i,j)\Big] = 0\ \ .
\eqno({\rm A}.21{\rm b})
$$

We now turn to the commutation relations of the
supersymmetry generators (A.2) with the vertical
nearest neighbor interactions $G^{\cal V}(i,j)$. To
simplify the calculations, we first notice that all
the braiding properties discussed so far, can be
translated into those appropriate for the vertical
anyons by simply using the angles $\Theta_{\cal V}(i,j)$
and ${\tilde \Theta}_{\cal V}(i,j)$ in place of
$\Theta_{\cal H}(i,j)$ and ${\tilde \Theta}_{\cal H}(i,j)$
respectively. Notice that $\Theta_{\cal V}(i,j)$ is defined
as $\Theta_{\cal H}(i,j)$ but rotated counterclockwise of
$\pi/2$ (compare Fig.1 and Fig.2). Similarly,
${\tilde \Theta}_{\cal V}(i,j)$ is defined as
${\tilde \Theta}_{\cal H}(i,j)$ but rotated of $\pi/2$.
Then, it is clear that all the manipulations we did before
naturally apply also to the vertical anyons (both with and
without the tilde sign). Since on the physical states of
the $t$-$J$ model the supercharges constructed by twiddled
vertical operators simply differ from those in (A.2) by a
global phase factor, we conclude that
$$
\Big[Q_\alpha\,,\,G^{\cal V}(i,j)\Big]
= \Big[Q^\dagger_\alpha\,,\,
G^{\cal H}(i,j)\Big] = 0\ \ .
\eqno({\rm A}.22)
$$
Finally, since the constraint $\phi(i)$ commutes
with the supercharges, from (A.21) and (A.22) we deduce
that the whole Hamiltonian of the $t$-$J$ model in the
slave anyon representation given in (4.24) is supersymmetric.

If we compare Eqs. (A.21) and (A.22) with Eqs. (4.28)
and (4.29), we see that using the representation (A.2)
for the supercharges, the quadratic interactions
${\hat G}^{\cal H}(i,j)$ and ${\hat G}^{\cal V}(i,j)$
are themselves supersymmetric, whereas if the
supercharges are written in the representation
(4.14) only the quartic interactions $\big|
{\hat G}^{\cal H}(i,j)\big|^2$ and
$\big|{\hat G}^{\cal V}(i,j)\big|^2$ are
invariant. In both cases, however, it is crucial
to use the constraint $\phi(i)=0$ which selects the
 physical states of the $t$-$J$ model. Clearly, in
order to check the invariance of the Hamiltonian (4.24),
the conditions expressed by (A.21) and (A.22) are sufficient
but not necessary. However, in view of a possible mean
field analysis of the slave anyon representation of the
$t$-$J$ model after a Hubbard-Stratonovich transformation,
it is more convenient to focus on the quadratic
interactions ${\hat G}^{\cal H}(i,j)$ and
${\hat G}^{\cal V}(i,j)$ instead of the quartic
Hamiltoinian which derives from them.
\vfill
\eject

\centerline{\bf References}
\vskip 0.7cm
\item{[1]} For a review see for example A.P.
Balachandran, E. Ercolessi, G. Morandi and A.M.
Srivastava, {\it Hubbard Model and Anyon
Superconductivity} (World Scientific Publishing
Co., Singapore 1990) and references therein;
\medskip
\item{[2]} P.W. Anderson, {\it Science}
{\bf 235} (1987) 1196;
\medskip
\item{[3]} P.W. Anderson, in {\it Frontiers and
Borderlines in Many-Particle Physics}, edited
by R.A. Broglia {\it et al.} (North-Holland,
Amsterdam, The Netherlands 1988);
\medskip
\item{[4]} F.C. Zhang and T.M. Rice, {\it Phys.
Rev.} {\bf B37} (1988) 3754;
\medskip
\item{[5]} P.B. Wiegmann, {\it Phys. Rev.
Lett.} {\bf 60} (1988) 821;
\medskip
\item{[6]} D. F\"orster, {\it Phys. Rev.
Lett.} {\bf 63} (1989) 2140; {\it Z. Phys.}
{\bf B82} (1991) 329;
\medskip
\item{[7]} P. Schlottmann, {\it Phys. Rev.}
{\bf B36} (1987) 5177; P.A. Bares, G.
Blatter and M. Ogata, {\it Phys. Rev.}
{\bf B44} (1991) 130;
\medskip
\item{[8]} F.H.L. Essler and V.E. Korepin, {\it Phys. Rev.}
{\bf B46} (1992) 9147;
\medskip
\item{[9]} C.K. Lai, {\it J. Math. Phys.} {\bf 15}
(1974) 1675; B. Sutherland, {\it Phys. Rev.}
{\bf B12} (1975) 3795;
\medskip
\item{[10]} A. Foerster and Karowski,
{\it Nucl. Phys.} {\bf B396} (1993) 611;
\medskip
\item{[11]} G. Baskaran, Z. Zou and P.W.
Anderson, {\it Solid State  Comm.}
{\bf 63} (1987) 973; Z. Zou and P.W.
Anderson, {\it Phys. Rev.} {\bf B37} (1988) 5594;
\medskip
\item{[12]}S.J. Barns, {\it J. Phys.} {\bf F6}
(1976) 1375; {\it J. Phys.} {\bf F7} (1977) 2637;
\medskip
\item{[13]} D. Yoshioka, {\it J. Phys. Soc. Jpn.}
{\bf 58} (1989) 1516;
\medskip
\item{[14]}F. Wilczek, in {\it Fractional Statistics
and Anyon Superconductivity} edited by F. Wilczek
(World Scientific Publishing Co., Singapore 1990);
\medskip
\item{[15]}A. Lerda, {\it Anyons: Quantum Mechanics of Particles
with Fractional Statistics} (Springer-Verlag,
Berlin, Germany 1992);
\medskip
\item{[16]} R.B. Laughlin, {\it Science} {\bf 242}
(1988) 525;
\medskip
\item{[17]} J. Fr\"ohlich, T. Kerler and P.A. Marchetti, {\it
Nucl. Phys.} {\bf B374} (1992) 511;
\medskip
\item{[18]} J. Fr\"ohlich and P.A. Marchetti,
{\it Phys.
Rev.} {\bf B46} (1992) 6535;
\medskip
\item{[19]} M. Gutzwiller, {\it Phys. Rev.} {\bf A137}
(1965) 1726;
\medskip
\item{[20]} D. Yoshioka, {\it J. Phys. Soc. Jpn.}
{\bf 58} (1989) 32;
\medskip
\item{[21]} I. Affleck and J.B. Marston, {\it Phys. Rev.}
{\bf B37} (1988) 3774; L.B. Ioffe and A.I. Larkin,
{\it Phys. Rev.} {\bf B39} (1990) 8988;
\medskip
\item{[22]} J.M. Leinaas and J. Myrheim,
{\it Nuovo Cim.}
{\bf 37B} (1977) 1; F. Wilczek, {\it Phys. Rev. Lett.}
{\bf 48} (1982) 114;
\medskip
\item{[23]} W. Siegel, {\it Nucl. Phys.}
{\bf B156} (1979) 135; S. Deser, R. Jackiw
and S. Templeton, {\it Ann. Phys.} {\bf 140} (1982) 372;
\medskip
\item{[24]} A.M. Polyakov, {\it Mod. Phys. Lett.}
{\bf A3} (1988) 325;
\medskip
\item{[25]} E. Witten, {\it Comm. Math. Phys.}
{\bf 121} (1989) 351;
\medskip
\item{[26]}M. L\"{u}scher, {\it Nucl. Phys.}
{\bf B326} (1989) 557; V.F. M\"{u}ller,
{\it Z. Phys.}
{\bf C47} (1990) 301;
\medskip
\item{[27]}D. Eliezer and G.W. Semenoff,
{\it Phys. Lett.} {\bf 266B} (1991) 375;
D. Eliezer, G.W. Semenoff and S.S.C. Wu,
{\it Mod. Phys. Lett.} {\bf A7} (1992) 513;
D. Eliezer and G.W. Semenoff,
{\it Ann. Phys.} {\bf 217} (1992) 66;
\medskip
\item{[28]} A. Lerda and S. Sciuto, Stony
Brook preprint ITP-SB-92-72, hep-th 9301100;
to be published on {\it Nucl. Phys.} {\bf B};
\medskip
\item{[29]} J. Fr\"{o}hlich and P.A. Marchetti,
{\it Comm. Math. Phys.} {\bf 112} (1987) 343;
\medskip
\item{[30]} S. Sarkar, {\it J. Phys.}
{\bf A23} (1990) L409; {\it J. Phys.} {\bf A24} (1991) 1137;
\medskip
\item{[31]} F.H.L. Essler, V.E. Korepin and
K. Schoutens, {\it Phys. Rev. Lett.} {\bf 68}
(1992) 2960; {\it Phys. Rev. Lett.} {\bf 70} (1993) 73;
\medskip
\item{[32]} J. Hubbard, {\it Phys. Rev. Lett.}
{\bf 3} (1959) 77; R.L. Stratonovich, {\it Sov.
 Phys. Doklady} {\bf 2} (1958) 416;
\medskip
\item{[33]} N. Nagaosa and P.A. Lee, {\it Phys.
Rev. Lett.} {\bf 64} (1990) 2450;
 P.B. Wiegmann, {\it Phys. Rev. Lett.} {\bf 65}
(1990) 2070;
\medskip
\item{[34]}P. Jordan and E.P. Wigner, {\it Z. Phys.}
{\bf 47} (1928) 631;
\medskip
\item{[35]} E. Fradkin and L.P. Kadanoff,
{\it Nucl. Phys.} {\bf B170} (1981) 1; E.
Fradkin, {\it Phys. Rev. Lett.}
{\bf 63} (1989) 322;
see also E. Fradkin, {\it Field Theories of
Condensed Matter Systems} (Addison-Wesley, Reading,
MA, USA 1991);
\medskip
\item{[36]}R. Caracciolo and M. R.- Monteiro,
{\it Phys. Lett.} {\bf 308B} (1993) 58; M. Frau, M.
R.-Monteiro and S. Sciuto, Torino preprint DFTT-16/93,
hep-th 9304108.
\vfill \eject

\centerline{\bf Figure Captions}
\vskip 1cm
\item{Fig. 1}{Representation of the lattice angle
$\Theta_{\cal H}(i,j)$
between the base point $B$ and $i$ measured
along the horizontal curve ${\cal C}_{\cal H}(i)$
from the point
$j^*$.}
\bigskip
\item{Fig. 2}{Representation of the lattice
angle
$\Theta_{\cal V}(i,j)$
between the base point $B'$ and $i$ measured
along the vertical curve ${\cal C}_{\cal V}(i)$
from the point
${}^*j$.}
\bigskip
\item{Fig. 3}{Representation of the lattice angle
${\tilde \Theta}_{\cal H}(i,j)$
between the base point $B$ and $i$
measured
along the curve ${\cal C}_{\cal H}(i)$ from the point
${}_*j$.}

\vfill
\eject
\nopagenumbers
\hskip 9cm \vbox{\hbox{DFTT 30/93}
\hbox{ITP-SB-93-30}
\hbox{June 1993}}
\vskip 1.5cm
\centerline{{\bf SLAVE ANYONS IN THE $t$-$J$ MODEL}}
\centerline{{\bf AT THE SUPERSYMMETRIC POINT}~
\footnote{$^*$}{
Work supported in part by Ministero dell' Universit\`a e
della Ricerca
Scientifica e Tecnologica, and by NSF grant PHY 90-08936.}}
\vskip 0.6cm
\centerline{{\bf Alberto Lerda} ~~and~~ {\bf Stefano Sciuto}}
\vskip 0.3cm
\centerline{\sl Institute for Theoretical Physics, S.U.N.Y.
at Stony Brook}
\centerline{\sl Stony Brook, N.Y. 11794, U.S.A.}
\centerline{ and }
\centerline{\sl Dipartimento di Fisica Teorica}
\centerline{\sl Universit\'a di Torino, and I.N.F.N.
Sezione di Torino}
\centerline{\sl Via P. Giuria 1, I-10125 Torino, Italy}
\vskip 2.5cm
\centerline{{\bf Abstract}}
\vskip 0.7cm
\noindent
We discuss the properties of the
supersymmetric $t$-$J$ model in the formalism
of the slave operators. In particular we introduce
a generalized abelian bosonization for the model in
two dimensions, and show that holons and spinons can
be anyons of arbitrary complementary statistics (slave
anyon representation). The braiding properties of these
anyonic operators are thoroughly analyzed, and are used
to provide an explicit linear realization of the
superalgebra $SU(1|2)$. Finally, we prove that the
Hamiltonian of the $t$-$J$ model in the slave anyon
representation is invariant under $SU(1|2)$ for $J=2\,t$.

\bye
\end